\documentclass{aa}
\newcommand{\varcsec}{^{\prime\prime}}

\newcommand\ionn[2]{#1$\,${\scshape{#2}}}%

\usepackage{amssymb}
\usepackage{graphicx}
\usepackage[bookmarks = true,colorlinks = true,linkcolor = blue,pdftex,urlcolor = blue,citecolor = blue,bookmarks = true,linktocpage = true,hyperindex = true,breaklinks]{hyperref}
\usepackage{microtype}
\usepackage{color} 
\usepackage{xfrac} 
\usepackage{inputenc}
\usepackage{fontenc}
\usepackage{natbib}

\bibpunct{(}{)}{;}{a}{}{,}

\usepackage{txfonts}
\begin{document}
   \title{Magnetic field variations associated with umbral flashes and penumbral waves}
   
   \subtitle{}

  \author{Jayant Joshi\inst{\ref{ISP},\ref{ROC},\ref{ITA}}
           \and Jaime de la Cruz Rodr\'{\i}guez\inst{\ref{ISP}}}           
   \institute{   Institute for Solar Physics, Department of Astronomy, 
                      Stockholm University, AlbaNova University Centre,
                      SE-106 91 Stockholm, Sweden, \label{ISP} 
                      \and Rosseland Centre for Solar Physics, 
                      University of Oslo, P.O. Box 1029 Blindern, NO-0315 Oslo, Norway \label{ROC} 
                      \and Institute of Theoretical Astrophysics, 
                      University of Oslo, P.O. Box 1029 Blindern, NO-0315 Oslo, Norway \label{ITA}\\
                      \email{jayant.joshi@astro.uio.no}
                 }
   \date{Received; accepted}

   \titlerunning{magnetic field variations in umbral flashes and penumbral waves}
   \authorrunning{Jayant Joshi et. al.}


\abstract
{Umbral flashes (UF) and running penumbral waves (RPWs) in sunspot chromospheres leave a dramatic
 imprint in the intensity profile of the \ion{Ca}{ii}~8542\,\AA\ line. Recent studies have focussed on
 also explaining the observed polarization profiles, that show even more dramatic variations during 
 the passage of these shock fronts. While most of these variations can be exaplined with an almost 
 constant magnetic field as a function of time, several studies have reported changes in the inferred
 magnetic field strenght during UF phases. These changes could be explained by opacity effects or
 by intrinsic changes in the magnetic field strength.}
{In this study we investigate the origin of these periodic variations of the magnetic field strength
by analyzing a time-series of high temporal cadence observations acquired in the \ion{Ca}{ii}~8542\,\AA\, line 
with the CRISP instrument at the Swedish 1-m Solar Telescope. In particular, we analyze how the inferred
geometrical height scale changes between quiescent and UF phases, and whether those changes are 
enough to explain the observed changes in $B$. }   
{We have performed non-LTE data inversions with the NICOLE code of a time-series of very 
high spatio-temporal resolution observations in the \ion{Ca}{ii}~8542\,\AA\, and  \ion{Fe}{i}~6301.5 \&
6302.5\,\AA\, lines. We analyze in detail the variations of the different physical parameters of the model
as a function of time.}
{Our results indicate that the \ion{Ca}{ii}~8542\,\AA\, line in sunspots is greatly sensitive to magnetic fields
at $\log\tau_{500}=-5$ (hereafter $\log\tau=-5$) during UFs and quiescence. However this optical depth value
does not correspond to the same geometrical height during the two phases. Our results indicate that during UFs 
and RPWs the $\log\tau=-5$ is located at a higher geometrical height than during quiescence. Additionally, 
the inferred magnetic field values are higher in UFs (up to $\sim270$~G) and in RPWs ($\sim100$~G).
}
{Our results suggest that opacity changes caused by UFs and RPWs cannot explain the
  observed temporal variations in the magnetic field, as the line seems to form at higher geometrical
  heights where the field is expected to be lower.}

\keywords{Sun: magnetic field - Sun: activity - Sun: chromosphere - Techniques: polarimetric} 

\maketitle
%
  
\section{Introduction} \label{sec_1}

Oscillations in sunspots have been studied in the outer layers of the Sun, 
preferentially using intensity or Doppler measurements from different spectral diagnostics.

In the chromosphere the imprint of these waves is particularly dramatic as they become shocks due to the 
steep decrease of density between the photosphere and chromosphere
 \citep{1984ApJ...277..874L,1986ApJ...301..992L,2007ApJ...671.1005B}. \citet{1969SoPh....7..351B} 
and \citet{1969SoPh....7..366W} identified for the first time sudden intensity enhancements in the core of the 
\ionn{Ca}{ii}\,K line in sunspot umbrae (commonly known as umbral flashes), although most 
chromospheric lines show a similar behavior. A sawtooth pattern of an upward propagating 
shock wave in umbrae are observed using different chromospheric and transition region lines 
\citep[e.g.,][]{2003A&A...403..277R, 2006ApJ...640.1153C, 2013A&A...556A.115D,2014ApJ...786..137T,2016ApJ...831...24K}.

The dominant period of these oscillations in umbrae is three minute in the chromosphere as well 
as in the transition region and corona \citep{1982ApJ...253..939G,1987SoPh..108...61G,1987ApJ...312..457T,1995A&A...300..539K,
2002A&A...387L..13D,2003A&A...403..277R,2006ApJ...640.1153C,2014ApJ...786..137T}. 
Running penumbral waves (RPWs) are produced by a similar physical process than umbral flashes, 
shock waves propagating along a magnetized atmosphere, but in this case the vertical propagation of the shock is slower 
than in the umbra because the magnetic field becomes more horizontal in the penumbra and the shock propagates
along those field lines \citep[e.g.,][]{2007ApJ...671.1005B}. 

\begin{figure*}
\centering
       \includegraphics[width = \textwidth]{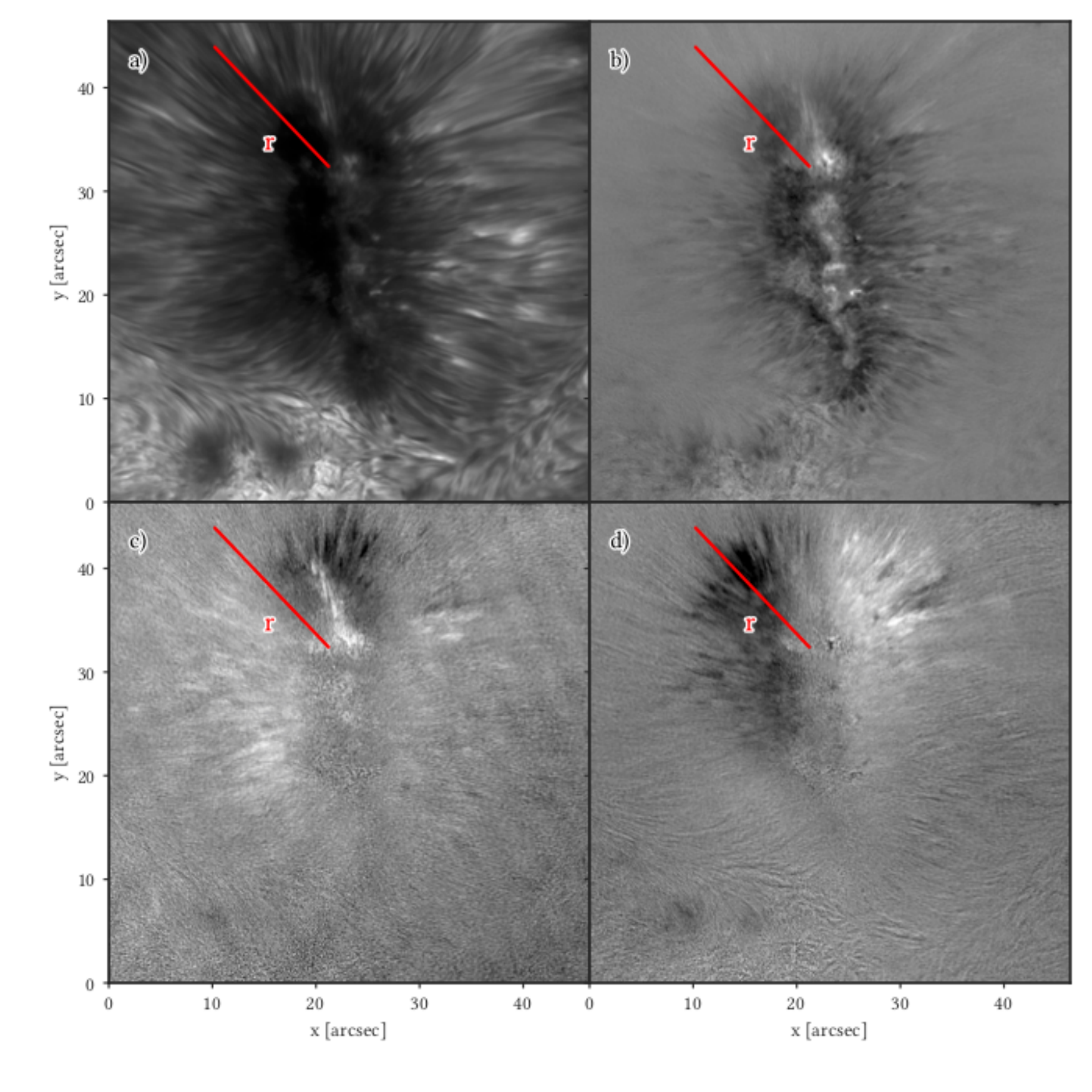}
       \caption{Field of view comprising the leading sunspot in the active region NOAA 11793 
               observed on 22 July 2013. Panel \textbf{a)} displays image obtained at the line center of the \ionn{Ca}{ii}\,8542\,\AA\, 
               spectral line using CRISP. Panels \textbf{b)}, \textbf{c)} and \textbf{d)} depict Stokes $V/I_{\rm{c}}$, $Q/I_{\rm{c}}$, 
                and, $U/I_{\rm{c}}$ maps at --350~m\AA\, from the line center.
                The $red$ cut marked as $r$ indicate the position for which the temporal evolution of the atmospheric
                parameters is analyzed in this study.
               }               
               \label{context}       
\end{figure*}


Although magnetic fields have been studied in sunspot umbra \citep{2000Sci...288.1396S}, only recently
\citet{2013A&A...556A.115D} analyzed magnetic field oscillations in the chromosphere. The latter found
that the magnetic field oscillates with an amplitude of $\sim$200\,G in the sunspot penumbra due to RPWs.
However, \citet{2013A&A...556A.115D} did not find significant variations of the magnetic field in the 
umbra. \citet{2017ApJ...845..102H} also studied umbral flashes in a sunspot chromosphere 
and they found that the strength of the vertical component of the magnetic field is slightly reduced 
during the flash phase compared to that in the quiescent phase.

Several authors have studied magnetic field oscillations in the photosphere, but these studies do not
present a consistent picture and often contradict each-other. For example,
variations in the magnetic field with different amplitudes, from zero to few tens of Gauss has been 
reported in a number of studies \citep[see e.g.][]{1997AN....318..129L,1997SoPh..172...69H,1998ApJ...497..464L,
1998A&A...335L..97R,2000SoPh..191...97K,1990SoPh..125...31B,2000ApJ...534..989B,2003SoPh..218...85B, 2012SoPh..280..347K}. 

Some authors, for example, \citet{1999ASSL..243..337R}, \citet{2000ApJ...534..989B}, 
\citet{2003A&A...410.1023R}, \citet{2003ApJ...588..606K}, and, \citet{2015LRSP...12....6K} have suggested
that the observed temporal variations of photospheric magnetic fields in sunspots could be an opacity 
effect that changes the effective formation height of the line due to oscillations in the thermodynamical
parameters. According to this idea, the vertical magnetic field gradient of a sunspot, in combination with
oscillating formation height of the spectral line under consideration can lead to false observations of
oscillation in the magnetic field.  Similarly, in the chromosphere, \citet{2017ApJ...845..102H} speculate 
that their results could be compatible with enhancements in the opacity of the \ionn{Ca}{ii}\,8542\,\AA\,
line during the flash phase. 

In this paper we study the evolution of physical parameters in a sunspot chromosphere during UFs, focusing 
our analysis in temporal fluctuations of the derived magnetic field vector in the photosphere and chromosphere.
We explore the origin of magnetic field oscillations in a sunspot and their relation to opacity
changes in the \ionn{Ca}{ii}\,8542\,\AA\, line.
\begin{figure*}
\centering
       \includegraphics[width = 0.96\textwidth]{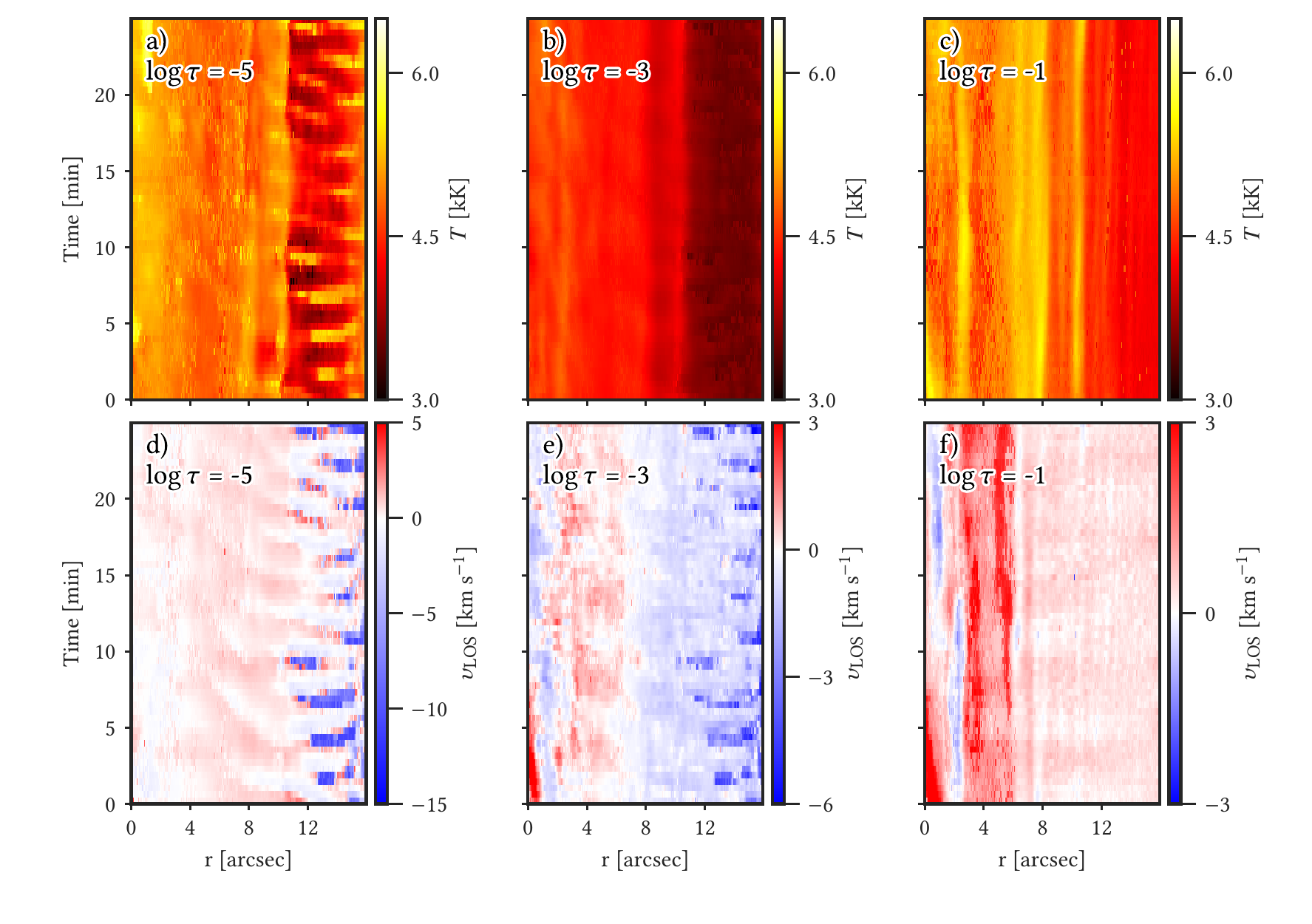}
       \caption{Temporal evolution of the temperature (up) and line-of-sight velocity (bottom) at
       	 	$\log \tau = -5, -3$, and, $-1$. Panels \textbf{a)}, \textbf{b)}, \textbf{d)} and \textbf{e)} illustrate the results from 
       	 	our inversions of the \ion{Ca}{ii}~8542\,\AA\, line, whereas panels \textbf{c)} and \textbf{f)} are taken from 
       	 	the inversion of the \ion{Fe}{i}~6301.5 \& 6302.5\,\AA lines.  $r=0$ corresponds 
             to the outer most point from the sunspot in the $red$ cut marked in Fig.~\ref{context}.}
       \label{par_map}
       
\end{figure*}
\section{Observations} \label{sec_2}  

Our observations were recorded in full-Stokes mode on the 22 July 2013 starting at 08:33~UT using the CRISP 
spectropolarimeter \citep{ 2006A&A...447.1111S, 2008ApJ...689L..69S} at the Swedish 1-m Solar Telescope
\citep[SST,][]{2003SPIE.4853..341S}. The observed field of view (FOV) consists of the leading sunspot in active region NOAA~11793.

The observations were acquired in the \ionn{Ca}{ii}~8542~\AA\ line 
at 21 wavelength points that sample a range of $\pm 1.750$~\AA\ from line center in an irregular grid of line positions. Close to the
line center, the line positions are sampled with a step of 70~m\AA\, and that grid becomes increasingly
more sparse in the broad photospheric wings of this line. We also acquired co-temporal data in the 
\ionn{Fe}{i}~6301.5 \& 6302.5~\AA\ lines with 18 wavelength points. The 
\ionn{Fe}{i}~6301.5 \& 6302.5~\AA\ lines were observed with a sampling of 40~m\AA\, close to line centers and 
with relatively sparse sampling in line wings. The total cadence of these observations is 25~s and 
the duration of the complete time series is 25~min.The data were reduced using the
CRISPRED pipeline \citep{2015A&A...573A..40D, 2012A&A...548A.114H, 2011A&A...534A..45S}, including 
Multi-Object-Multi-Frame-Blind-Deconvolution processing \citep[MOMFBD,][]{2005SoPh..228..191V} of 
the entire time series.

Fig.~\ref{context}-\textbf{a)} illustrates our target displayed in the line center of the \ionn{Ca}{ii}~8542~\AA\ line.
Maps of Stokes $V/I_{\rm{c}}$, $Q/I_{\rm{c}}$, and, $U/I_{\rm{c}}$ at a wavelength offset 
of $-350$~m\AA\ from the line center are displayed in panels \textbf{b)}, \textbf{c)} and \textbf{d)}, respectively. 
$I_{\rm{c}}$ represents average continuum intensity. The red slit marked on the FOV indicates the location of 
the spectra that we have extracted to perform our temporal analysis. The solar limb is located towards the 
upper part of the image.

%

\section{Inversions} \label{sec_3}

\begin{figure*}
\centering
       \includegraphics[width = 0.96\textwidth]{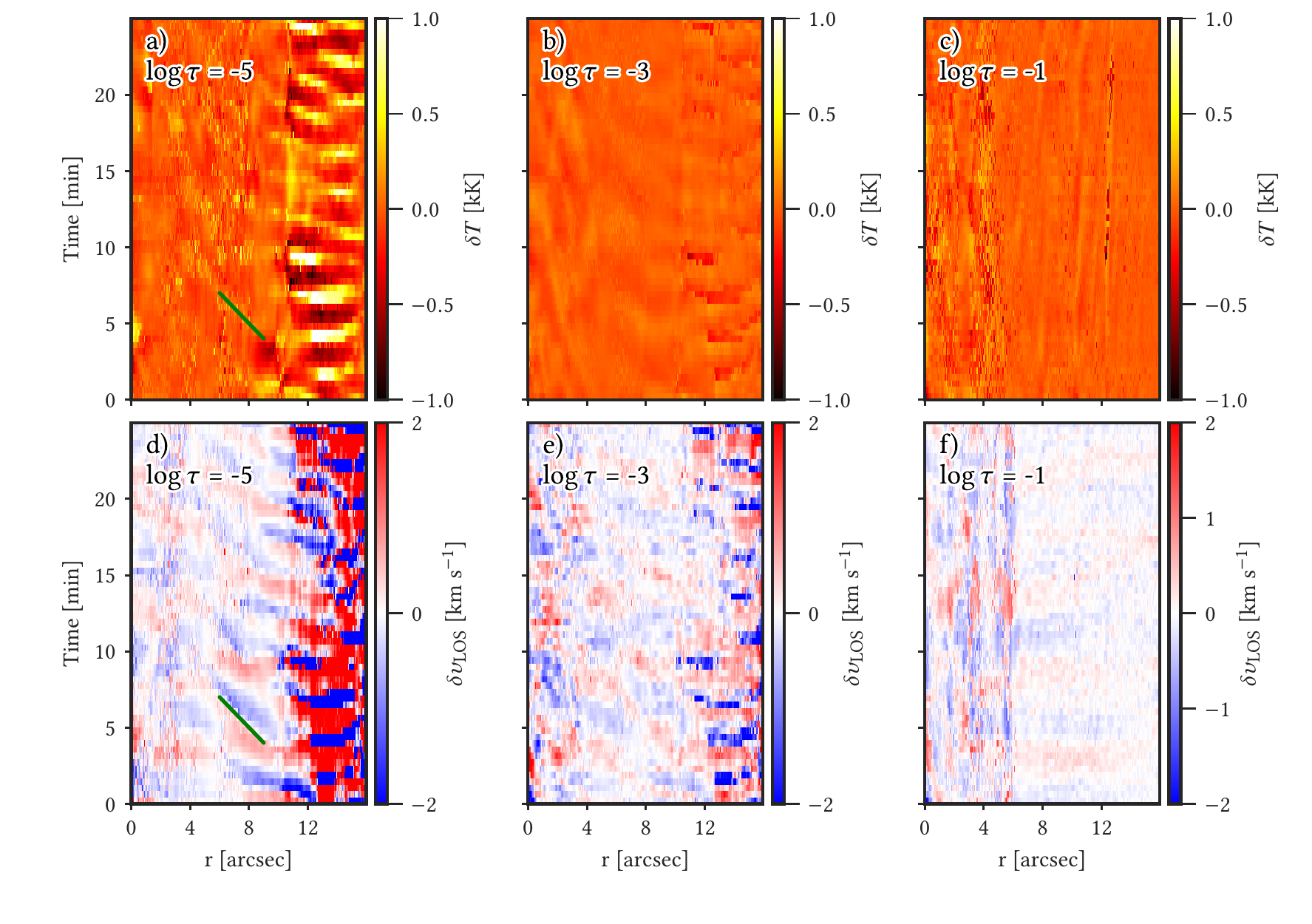}
       \caption{Same as Fig.~\ref{par_map}, but in this case we illustrate the residual variations, $\delta T$ and 
                $\delta\upsilon_{\rm{LOS}}$ after subtracting the local background. The \textit{green} line in panels \textbf{a)} and \textbf{d)} 
                traces a running penumbral wave (RPW).}
       \label{par_map_diff}
\end{figure*}

\begin{figure}
\centering
       \includegraphics[width = \columnwidth]{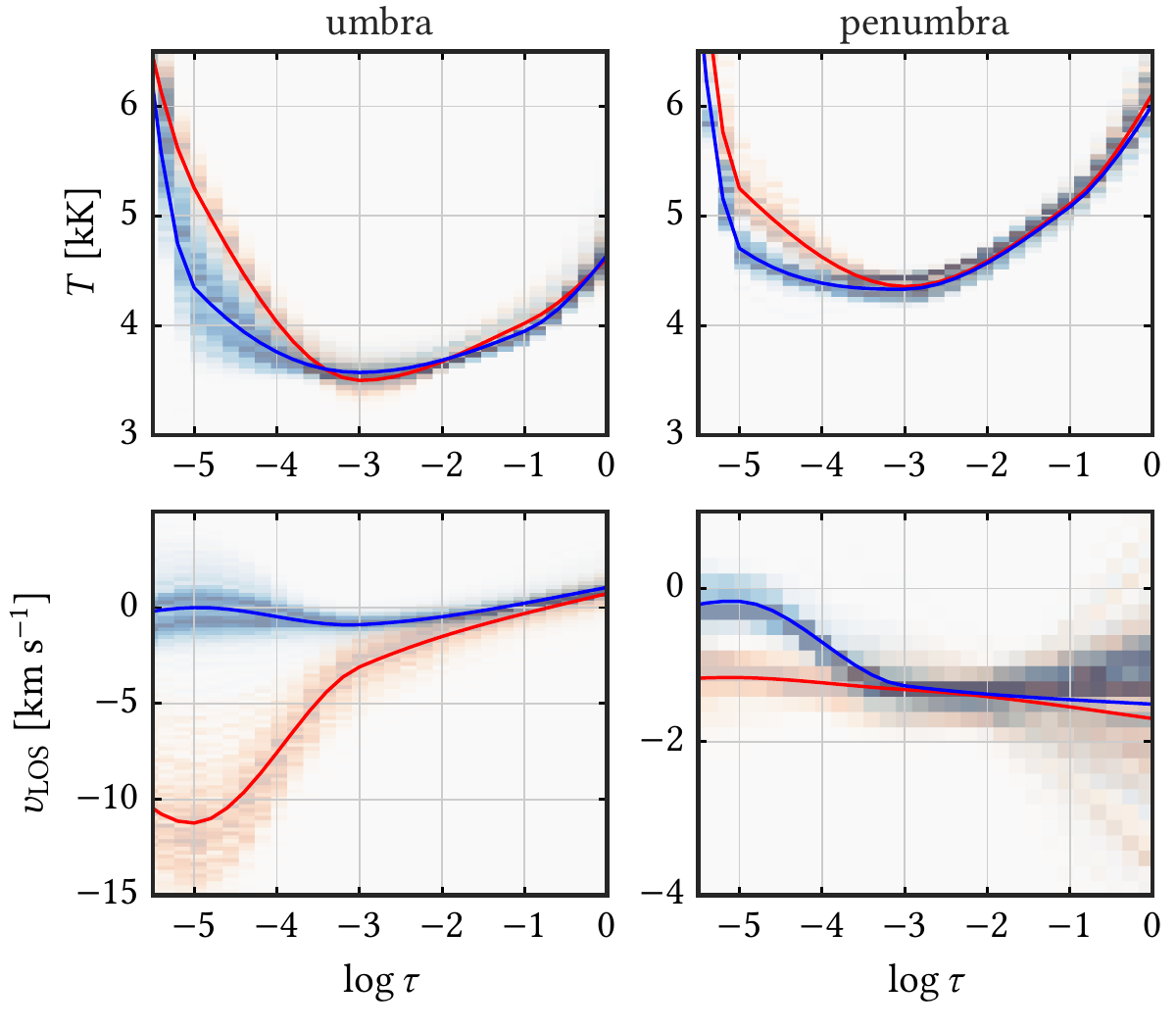}
       \caption{Averaged stratification of $T$ and $\upsilon_{\rm{LOS}}$  as a function of optical depth obtained from the inversions of 
               the  \ionn{Ca}{ii}\,8542\,\AA\, line. The \textit{Upper left} and \textit{lower left} panel show the stratification of $T$
               and $\upsilon_{\rm{LOS}}$ in the umbra. The \textit{Red} and \textit{blue} curves represent the umbral flash and
                quiescent umbra atmosphere respectively. The panels on the right
               similarly show the stratification of the same physical parameters in the penumbra.
                 The \textit{shaded} areas in all panels demonstrate the distribution of the 
               parameters normalized at each $\log \tau$ position.} 
       \label{avg_atm}
\end{figure}

We used the non-LTE inversion code NICOLE \citep{2015A&A...577A...7S} to obtain atmospheric parameters in the 
chromosphere and photosphere of our sunspot observations. The parameters of our model atmospheres are 
calculated in an optical-depth grid with 5 points per decade, which should suffice to compute very accurate 
intensities with DELO-Bezier formal solvers \citep[see][]{2013ApJ...764...33D, 2017ApJ...840..107J}. 
We refer to the code description paper for further details. We have included the effect of \ion{Ca}{ii} 
isotopic splitting in our inversions which can affect the derivation of Doppler velocities when 
it is not included in the inversion \citep[see][]{2014ApJ...784L..17L}.

We carried out the inversions only for a time series extracted along the red cut shown in Fig.~\ref{context}.
We performed the inversions separately for the \ionn{Ca}{ii}\,8542\,\AA\, and \ionn{Fe}{i}\,6301.5\,\&\,6302.5\,\AA\,
lines. For the \ionn{Ca}{ii}\,8542\,\AA\, line, we used five nodes located at $\log\tau$ = --7, --5, --3, --1 and, 1 
for temperature, $T$, four nodes at  $\log\tau$ = --7, --5, --3 and, 1 
for line-of-sight (LOS) velocity, $\upsilon_{\rm LOS}$. $\tau$ represents the continuum optical-depth at 500\,nm.
For all three component of the magnetic field, $B_x$, $B_y$, and, $B_z$ we used two nodes,
i.e., the magnetic field varies linearly with $\log \tau$.  $B_z$ is the LOS component of the magnetic 
field and $B_x$ and $B_y$ are the two component orthogonal to each-other and $B_z$. 
The micro-turbulence, $\upsilon_{\rm{turb}}$, is assumed to be constant as a function of optical-depth.

We have carried out test inversions to select the minimum number of nodes that allows to properly reproduce the
observed spectra in the umbra, penumbra and outside the Sunspot. In the penumbra, good fits to the observed Stokes 
profiles were achieved even when we only allowed linear gradient in $\upsilon_{\rm LOS}$ with respect to $\log\tau$. 
But UF atmospheres required more nodes in $\upsilon_{\rm LOS}$ to fit the observed profiles. Hence, the inversions 
are carried out in two cycles, in the first cycle we used only two nodes for $\upsilon_{\rm LOS}$ and single node for
$B_x$, $B_y$, and, $B_z$. In the second cycle, four nodes are used for $\upsilon_{\rm LOS}$ and two nodes for
$B_x$, $B_y$, and, $B_z$. The results from the first cycle are used as an input in the second cycle.

We have another set of inversions for the \ionn{Ca}{ii}\,8542\,\AA\, line where the model atmosphere is very similar as describe 
above, except the magnetic field is assumed to be constant with optical depth. This additional set of inversions allow us to
compare the magnetic field properties of the sunspot obtained with two different approaches of the inversions.  

\citet{2016MNRAS.459.3363Q} analyzed response functions, $RF\rm{s}$, of the \ionn{Ca}{ii}\,8542\,\AA\, line 
for different thermodynamical parameter and the magnetic field for the quiet-Sun FAL-C model \citep{1993ApJ...406..319F}.
Their analysis showed that the \ionn{Ca}{ii}\,8542\,\AA\, line has almost negligible or zero response
to $\upsilon_{\rm LOS}$ in the atmosphere below $\log\tau = -3$, and because of that we avoided adding a node for 
$\upsilon_{\rm LOS}$ between $\log\tau = -3$ and 1 in our inversions. However, $RF\rm{s}$ could be significantly 
different for a sunspot model compared to that for FAL-C model.   

For the photospheric magnetic field measurements, we inverted the \ionn{Fe}{i}\,6301.5\,\&\,6302.5\,\AA\, lines using 
four nodes, $\log\tau$ = --5, --3, --1 and, 1 for $T$ and two nodes for $\upsilon_{\rm LOS}$.
We assume constant $B_x$, $B_y$, $B_z$ and, $\upsilon_{\rm{turb}}$ as a function of optical depth. 

Since we are working with data that include the umbra-penumbra boundary, our data can be particularly affected by
straylight from residual (uncorrected) atmospheric aberrations \citep{2010A&A...521A..68S}. Additionally, the 
flat-fielding method for the 8542 data includes a backscatter correction step that can slightly modify the 
contrast of the image \citep[for more details see Appendix A1 of][]{2013A&A...556A.115D}. For these reasons,
and to avoid potentially controversial data deconvolutions, we decided to perform the inversion 
of the \ionn{Ca}{ii}\,8542\,\AA\, line separately from the \ionn{Fe}{i}\,6301.5\,\&\,6302.5\,\AA\, lines.

Hereafter, the chromospheric inversion results displayed at $\log\tau$ = --5 and --3 are taken from the 
inversions of the \ionn{Ca}{ii}\,8542\,\AA\, line, whereas the photospheric parameters shown at  $\log\tau = -1$ are 
obtained from the inversions of the \ionn{Fe}{i}\,6301.5\,\&\,6302.5\,\AA\, lines. 

\section{Results and analysis} \label{sec_4}

\subsection{Umbral flashes in temperature and LOS velocity}
The temporal evolution of $T$ and $\upsilon_{\rm{LOS}}$  are presented in Fig.~\ref{par_map}. 
At $\log\tau = -5$, $T$ and $\upsilon_{\rm{LOS}}$ reveal a clear oscillatory pattern in the umbra. 
At $\log\tau = -3$ the imprint of these oscillations is still visible but not as prominent 
as at $\log\tau = -5$. At $\log\tau = -1$, both quantities present spatial changes and a very 
smooth photospheric temporal evolution. 

In order to emphasize the smaller scale fluctuations that are present in the penumbra,
in Fig.~\ref{par_map_diff} we have extracted the local background at
each location by fitting a third order polynomial along the temporal dimension for each slit position.
Hereafter, these residual variations are denoted with a $\delta$ in front of the physical variable name.

The signature of RPWs is now clearly visible in $\delta T$ and $\delta \upsilon_{\rm{LOS}}$ at $\log\tau$ = --5 and --3.
We have traced a RPW (see the green line in panels  \textbf{a)} and \textbf{d)})
which shows an outward motion from the inner penumbra towards the outer penumbra with speed of $\sim$9\,km\,s$^{-1}$.

Fig.~\ref{par_map} and \ref{par_map_diff} show persistent variations in the temperature and LOS velocity during the passage of
UFs and RPW that are believed to be responsible for opacity changes in the \ion{Ca}{ii}~8542\,\AA\, line.

We calibrated the absolute reference $\upsilon_{\rm{LOS}}$ by assuming that the average photospheric velocity in the umbra is equal to zero. The 
\ionn{Ca}{ii}\,8542\,\AA\, line is not very sensitive to photospheric LOS velocities, so our calibration of $\upsilon_{\rm{LOS}}$ may be affected by large uncertainties, but it is the best we can do based on prior knowledge of the observed target.

\subsubsection{Average quiescent and flash atmospheres}
\emph{Umbra:}
The average temperature stratification in UFs and quiescence in the umbra are relatively 
similar up to $\log\tau = -3$ (see Fig.~\ref{avg_atm}).  Around $\log\tau = -5$, 
atmosphere in UFs can be on average up to 1\,kK\, hotter than the quiescent 
umbra as reported by \citet{2013A&A...556A.115D}. The LOS velocity shows an upflowing atmosphere
during the UF phase in the chromsphere, but a rather static situation during quiescence.
 Below $\log\tau = -3$, $\upsilon_{\rm{LOS}}$ is almost identical during UFs and in the
quiescent phase.

\emph{Penumbra}:
In the penumbra we find a similar behavior as in the umbra. At $\log\tau = -5$,  $T$ is
0.5~kK higher during the hotter phase of RPWs. Similarly, $\upsilon_{\rm{LOS}}$ at $\log\tau = -5$ 
is on average around --1.2\,km\,s$^{-1}$ during the hotter 
phase of RPWs, otherwise is it close to zero. 

We note that the average stratification of the parameters
shown in Fig.~\ref{avg_atm} are calculated from the inversion results of the \ionn{Ca}{ii}\,8542\,\AA\, 
line. The latter provides limited velocity diagnostic capabilities in the photosphere due to the very 
broad (photospheric) wings of the \ion{Ca}{ii} line.

\subsection{Time evolution of $B$}
In the present study, we have assessed the quality of our fits during UFs and RPWs. We have
also studied the sensitivity of the \ion{Ca}{ii}~8542 line to magnetic fields and magnetic field
gradients in sunspot atmospheres using response functions. These studies can be found in 
Appendixes~\ref{ap:fit} and \ref{ap:sen}. The very short summary of that study
is that in the derived sunspot atmospheres, the response function of the \ion{Ca}{ii}~8542 line peaks approximately 
 at $\log \tau = -5$.

\begin{figure}
\centering
       \includegraphics[width = \columnwidth]{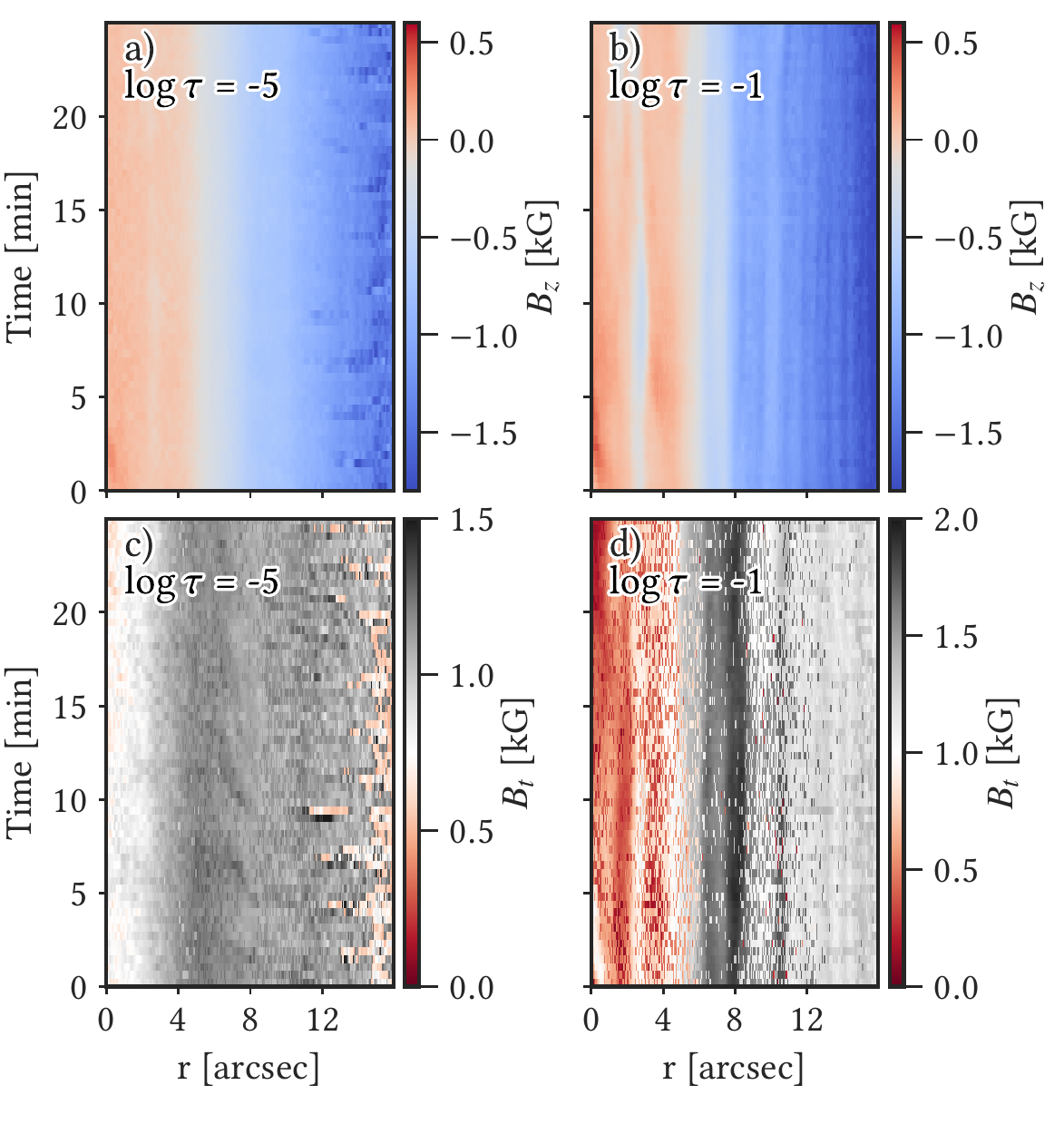}
       \caption{Temporal evolution of the magnetic field in the chromosphere and in the photosphere. Panels \textbf{a)} and \textbf{c)}
                depict the LOS component of the magnetic field, $B_z$, and the transverse component of
                the magnetic field, $B_t$, at $\log \tau = - 5.0$. $B_z$ and $B_t$ at 
                $\log \tau = - 1.0$ are shown in panels \textbf{b)} and \textbf{d)}.
                These results are solely obtained from the inversion of the \ionn{Ca}{ii}\,8542\,\AA\, line. }
       \label{par_map_mag}
\end{figure}

\begin{figure}
\centering
       \includegraphics[width = \columnwidth]{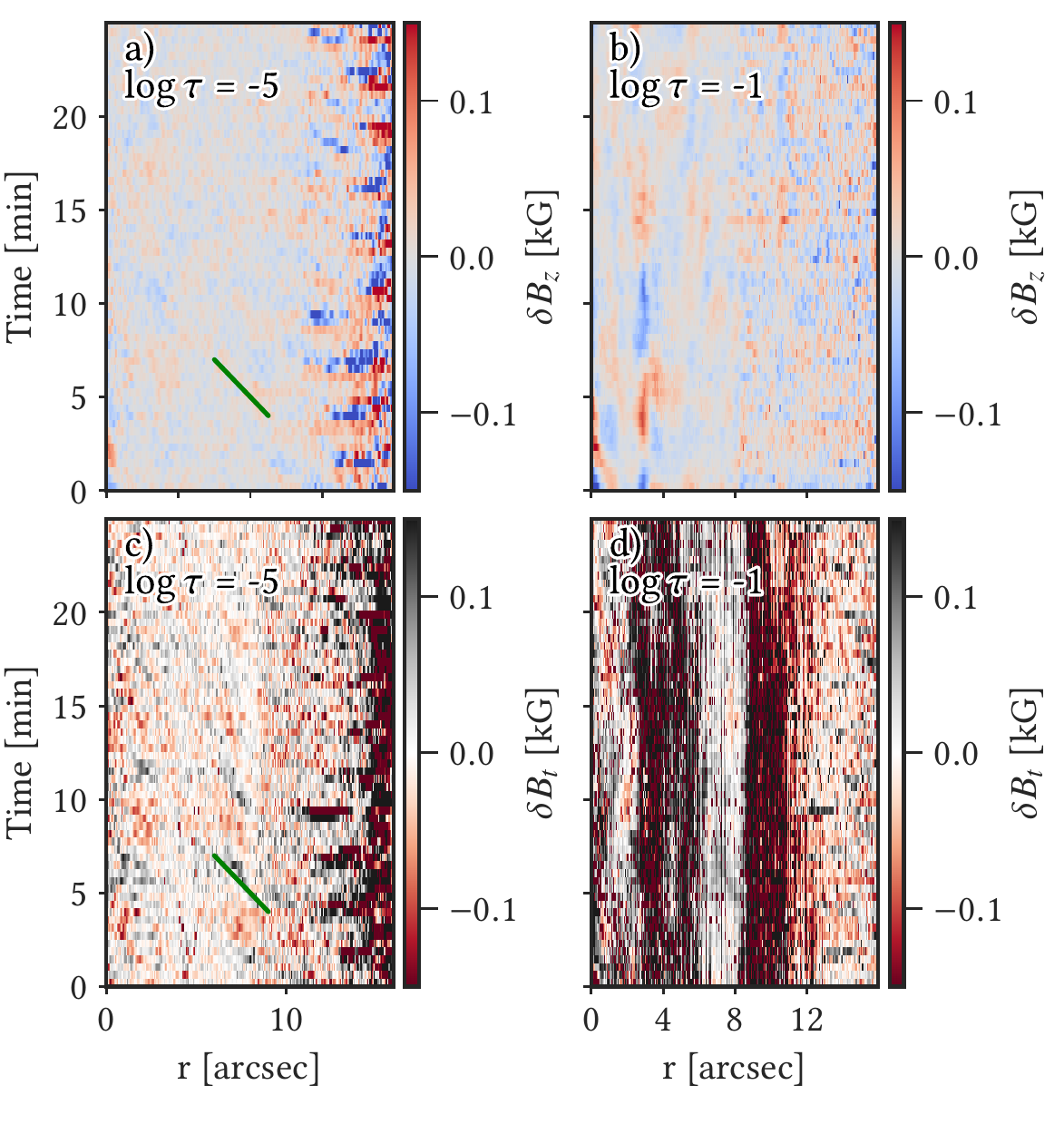}
       \caption{Same as Fig.~\ref{par_map_mag}, but in this case we show the residual variations, $\delta\mid B_z \mid$ 
       and $\delta B_t$ after removing the local background. The $green$ line in panels \textbf{a)} and \textbf{c)} represents the same RPW as in Fig.~\ref{par_map_diff}.}
       \label{par_map_mag_diff}
\end{figure}

Our magnetic field maps show the oscillatory imprint of UFs and RPWs as a function of time 
(see Fig.~\ref{par_map_mag} and Fig.~\ref{par_map_mag_diff} for the background compensated version). 
The amplitudes of these oscillations are larger in the umbra than in the penumbra, and they
have the same phase and period as those detected in $T$.
In the chromosphere, both components of the magnetic field show a very similar behavior,
except that $B_t$ at the very end of the slit (in the umbra) does not show the imprint of UF 
because there is no signal in $Q$ and $U$ in those pixels.
In the photosphere we detect a long period oscillation in $\delta B_z$ at $r<4\arcsec$, 
but otherwise there is no obvious evidence of UF or RPW in any of the physical parameters.

\begin{figure*}
\centering
       \includegraphics[width = 0.96\textwidth]{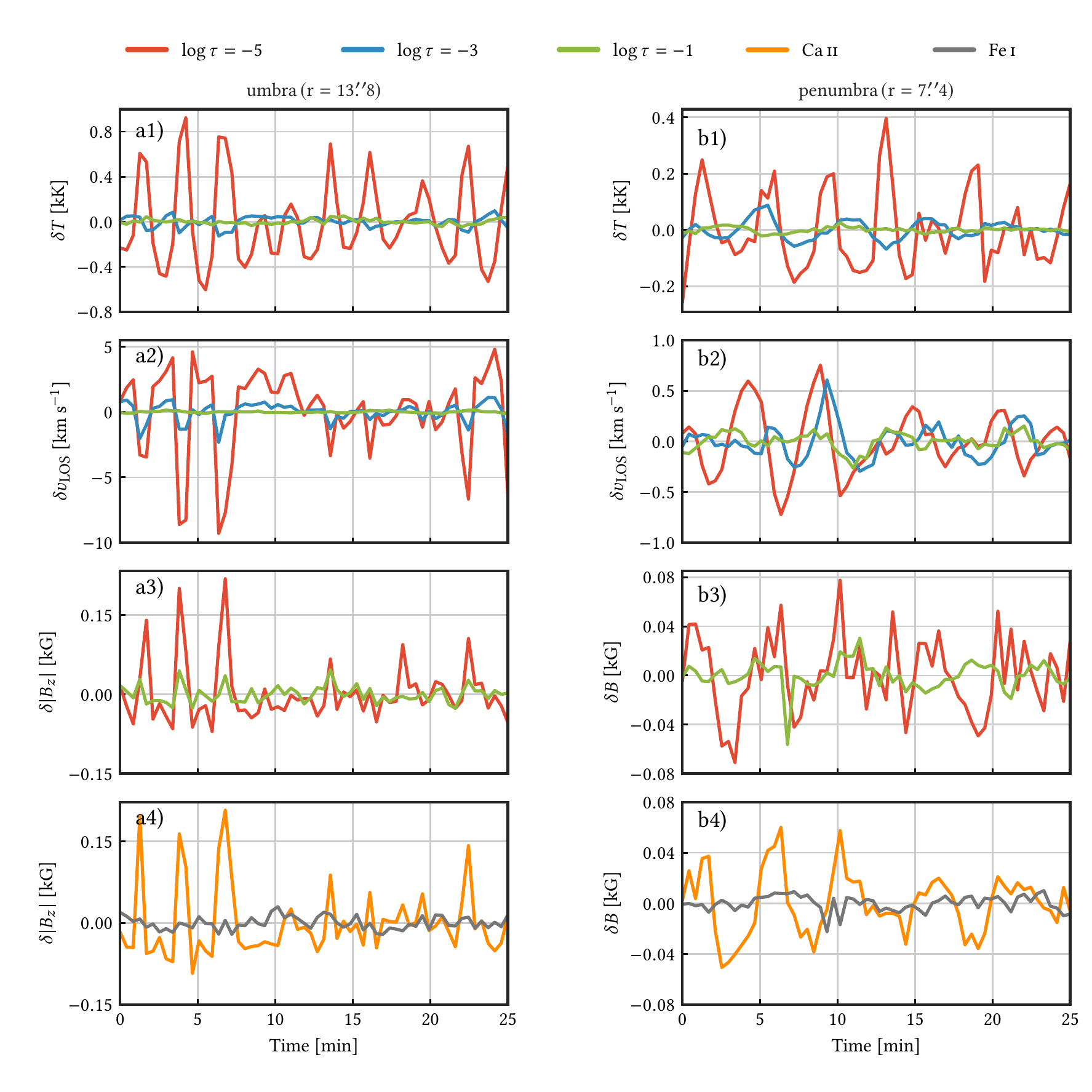}
       \caption{Temporal variations of the atmospheric parameters in the umbra 
               (\textit{left column}), and penumbra (\textit{right column}) after removing the local background. The curves plotted in these panels correspond to  
               $r = 13.\arcsec8$ and $r = 7.\arcsec4$ for the umbral and penumbral sets, respectively. From
               top to bottom, \textbf{a1)} - \textbf{a3)}: $\delta T$, $\delta \upsilon_{\rm{LOS}}$ and, 
               $\delta \mid B_z \mid$  from the inversions of the \ionn{Ca}{ii}\,8542\,\AA\, line including
                a linear gradients in the magnetic field as a function of $\log \tau$. \textbf{a4)}: $\delta \mid B_z \mid$ from the inversions of 
               the \ionn{Ca}{ii}\,8542\,\AA\, line  (orange curve) and the \ionn{Fe}{i} line pair at \,6302\,\AA\, (gray curve) 
               separately with the assumption that the magnetic field is depth independent. Panels
               \textbf{b1)} - \textbf{b4)} are similar to \textbf{a1)} - \textbf{a4)}, but the former depict 
                $\delta B$ instead of  $\delta \mid B_z \mid$. The \textit{red}, \textit{blue} and \textit{green} 
               curves correspond to $\log \tau = -5, -3$, and, $-1$.}     
               \label{par_var}
             \end{figure*}

We have selected two locations along the slit in the umbra and in the penumbra. 
Fig.~\ref{par_var} shows a comparison of all physical quantities at those locations.
This figure greatly illustrates an almost perfect correlation of all parameters in the umbra,
during UF passages. In the penumbra $\upsilon_{\rm{LOS}}$ seems to be slightly out
of phase with $T$ and $B$ during RPWs. But the correlation between $\delta T$ and $\delta |B_z|$ and $\delta B$
in both cases is remarkable and it seems to point to opacity effects. This figure also makes it very easy to appreciate how much
larger the chromospheric oscillations are when compared to those in the photosphere. For clarity, 
we have summarized in Table~\ref{tab:peak} some of the peak-to-peak values that are observed in Fig.~\ref{par_var}.

\begin{table}
 \centering
 \caption{Approximate peak-to-peak variations of the physical parameters at $\log\tau = -5$ shown in Fig.~\ref{par_var}.}
 \label{tab:peak}
 \begin{tabular}{l | c c c}
 \hline\hline
  \  & $\delta T$ [K] & $\delta\upsilon_\mathrm{LOS}$ [km s$^{-1}$]& $\delta |B_z|$ or $\delta |B|$ [G]\\
  \hline
  umbra & $500 - 1500$ & $5 - 12$ & $100 - 270$ \\
  penumbra & $200 - 500$ & $0.5 - 1.5$ & $50 - 100$ \\
\hline
 \end{tabular}

\end{table}    
 

\begin{figure}
\centering
       \includegraphics[width = \columnwidth]{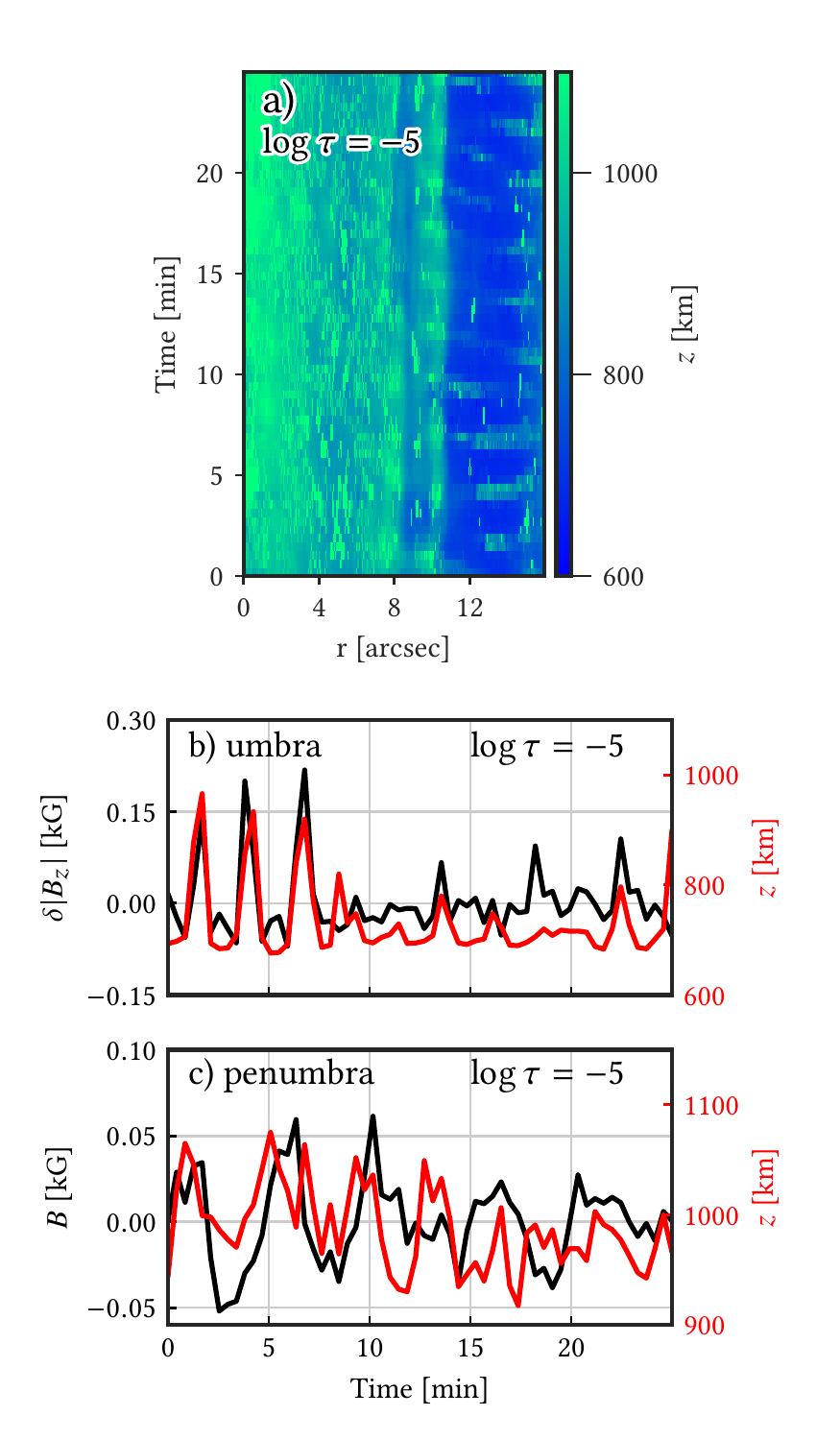}
       \caption{Space-time variation of $z(\log\tau = -5)$ (panel \textbf{a)}). Panel \textbf{b)} 
                represents the temporal variations of $\delta\mid B_z \mid$ and $z$ at $\log \tau = -5$ for an umbral pixel 
                ($r = 13.\arcsec8$) and panel \textbf{c)} depicts the temporal variation of $\delta B$ and $z$ at $\log \tau = -5$
                for a penumbral pixel ($r = 7.\arcsec4$).}  
                \label{map_z_tau}
\end{figure}

\begin{figure}
\centering
       \includegraphics[width = \columnwidth]{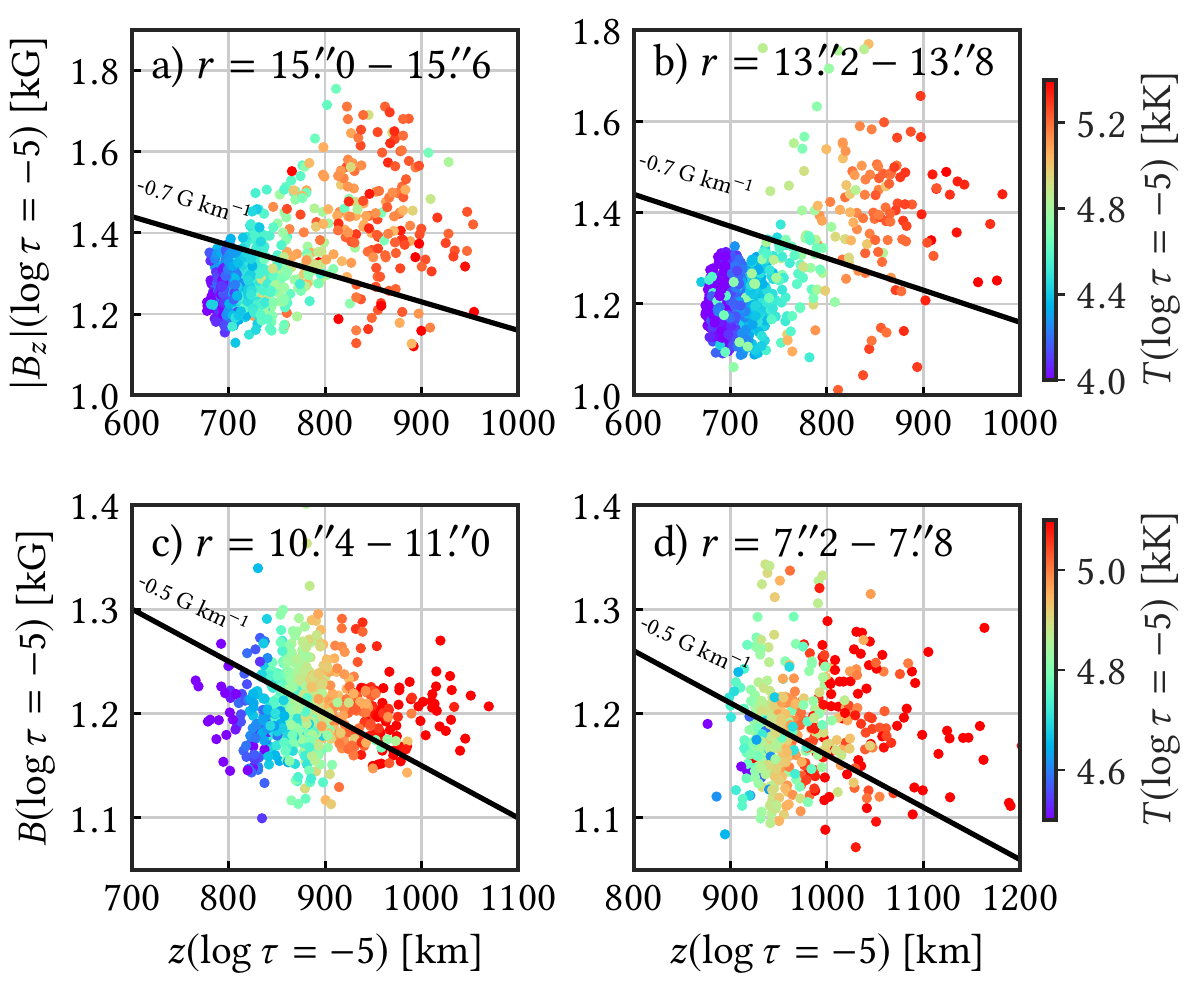}
       \caption{
       Distribution of $|B_z|$ as a function of $z(\log\tau=-5)$ for the time-series of two locations in the umbra (top) and two locations in the penumbra (bottom).
                Panel \textbf{a)}: Relation $\mid B_z \mid$ and  $z$ in the umbra at $r = 15.\arcsec0$ - $15.\arcsec6$. Panel \textbf{b)} is same 
                but for $r = 13.\arcsec2$ - $13.\arcsec8$. Panels \textbf{c)} and \textbf{d)} are taken at  $r = 10.\arcsec4$ - $11.\arcsec0$ and 
                $r = 7.\arcsec2$ - $7.\arcsec8$, respectively. The \textit{black} lines
                show the average vertical gradient of $\mid B_z \mid$ in the umbra (--0.7\,G\,km$^{-1}$) and penumbra (--0.5\,G\,km$^{-1}$). 
                The points have been color coded according to their temperature $T(\log \tau = -5)$.} \label{sct_z_b}
\end{figure}

\subsection{Opacity changes during the UFs and RPWs}
Both in the photosphere and chromosphere, several authors have suggested that the observed oscillations of 
the magnetic field in sunspots may be due to changes in line opacity induced by waves. The sunspot's 
magnetic field has a gradient with height and if the line opacity changes during the passage of UFs, then the core of this spectral line can sample
a different magnetic field regime. We analyze the changes in 
the geometrical height scale, $z$, as a consequence of changes in the thermodynamical parameters due to oscillations.  

We have computed the geometrical height that corresponds to $\log \tau = -5$ $[ \ z(\log \tau = -5) \ ]$,
assuming hydrostatic equilibrium. Although the latter is an approximation that may not reproduce
the exact geometrical scale of the Sun, it allows to perform a differential study of $z$ as a function of time, as illustrated in Fig.~\ref{map_z_tau}. Our results indicate that the magnetic field is amplified during UFs and RPWs, while the $z(\log \tau = -5)$ increases.

In the umbra, there is a clear variation of $z(\log \tau = -5)$ during UF
phases (up to $270$ km), which is tightly correlated with the variations displayed in the space-time maps of 
$T$, $\upsilon_{\rm{LOS}}$, $B_z$, and $B_t$.
In the penumbra this opacity effect has a somewhat smaller impact, and
the variations of $z(\log \tau = -5)$ are smaller ($\sim100$ km).

In order to show some more statistics, we have computed scatter plots in a portion of the umbra and penumbra (see Fig.~\ref{sct_z_b}). The color scaling represents the corresponding $T(\log\tau = -5)$.
The clusters of hot points (higher $T(\log\tau = -5)$) in panel \textbf{a)}, 
have an average value of $z(\log\tau = -5) \approx 850$\,km and $\mid B_z \mid(\log\tau = -5)$ equal to 1.45\,kG.
Cooler pixels are clustered around $z(\log\tau = -5) = 700$\,km and $\mid B_z \mid(\log\tau = -5) = 1.30$\,kG. 
A similar trend in the relation between $\mid B_z \mid(\log\tau = -5)$
and $z(\log\tau = -5)$ is present in panel \textbf{b)}, which corresponds to a different location along the slit also in the umbra.

This trend is completely different to what we expect from previous measurements of sunspot
magnetic fields as a function of height.
 We have derived the vertical gradient of the magnetic field between $\log \tau = -1$ and $\log \tau = -5$
using our approximate (hydrostatic equilibrium) $z$-scale. As we mentioned before, our inversions 
of $B_x$ and $B_y$ in the umbra are not very reliable. Therefore we have only calculated the vertical 
gradient of $B_z$ in the umbra. On average, the umbral vertical gradient of $B_z$ has a value around 
--0.7\, G\,km$^{-1}$, which is not the trend that we observe during the passage of shocks.

In the penumbra (panels \textbf{c)} and \textbf{d)}, there are no clear relations between $B(\log\tau = -5)$ and $z(\log\tau = -5)$. Although, there is a little hint of a trend 
suggesting that pixels with a higher value of $z(\log\tau = -5)$ harbor higher values of $B(\log\tau = -5)$. Intrinsically small temporal variations
of $B$, $z$ and $T$ in the penumbra might be the reason why no clear relation between these parameters is derived in 
the penumbra compared to that in the umbra. Moreover, the average vertical gradient of $B$ in the penumbra is approximately --0.5\, G\,km$^{-1}$. 

\section{Discussion and conclusion} \label{sec_5}

We have analyzed the temporal evolution of the magnetic field along a radial cut in the observed sunspot covering
the umbra and penumbra. In the umbra, the observed Stokes $Q$ and $U$ profiles in the \ionn{Ca}{ii}\,8542\,\AA\, line 
do not have sufficient signal, so only LOS magnetic field is measured with confidence. However, in the penumbra, 
all Stokes profiles have significant signal allowing the reliable measurement of all the magnetic field components.   

We have found that in the umbra the chromospheric LOS magnetic field, $B_{\rm{z}}(\log\tau = -5)$, varies periodically 
with a maximum variation in amplitude up to 270\,G. The period of these oscillations is 3\,min confirming the well
known 3 min oscillations in the chromosphere of sunspots. In the chromosphere, LOS magnetic field variations due
to UFs are correlated and anti-correlated to those in the temperature and LOS velocity, respectively. That means 
that during the passage of an UF, the chromospheric magnetic field increases along with the corresponding temperature
increase. 

Similarly, in the penumbra, the chromospheric magnetic field oscillates with an amplitude up to 100\,G. These 
variations in the magnetic field are correlated with a temperature rise due to RPWs. \citet{2013A&A...556A.115D} 
have reported fluctuations in the magnetic field of RPWs with an amplitude of 200\,G using the same spectral line 
and with similar observations as presented in the current paper. Although, \citet{2013A&A...556A.115D} did not find any 
temporal variations of the magnetic field in the umbra. Our results indicate no significant variations in the photospheric 
magnetic field of the sunspot.

We have also calculated the vertical gradient of the magnetic field between the photosphere and chromosphere of the
sunspot. The average vertical gradient of the LOS magnetic field in the umbra is --0.7\,G\,km$^{-1}$. In the penumbra, 
the magnetic field decreases with an average rate of --0.5\,G\,km$^{-1}$ in the vertical direction. The value of the 
estimated vertical gradient of the magnetic field is in agreement with those reported in the literature 
\citep[see, e.g.,][]{1995A&A...302..543R,2005ESASP.596E..59O,2015SoPh..290.1607S,2016A&A...596A...8J,2017A&A...604A..98J}.
These authors report the vertical gradient of the sunspot's magnetic field in the range of $-0.3$-$-1.0 \rm{~G~km}^{-1}$.

Our analysis of the magnetic field response function indicates that the \ionn{Ca}{ii}\,8542\,\AA\, line is most sensitive to
the magnetic field variations occurring at $\sim\log\tau = -5$ in the sunspot atmosphere, hence, the magnetic field 
measured at this depth is most reliable. We investigated the temporal variations in the geometrical height at $\log\tau = -5$ 
and its relation to the observed magnetic field variations in the chromosphere. The geometrical height is calculated 
assuming hydrostatic equilibrium. We are aware that the obtained geometrical height with the approximation of
hydrostatic equilibrium can be significantly different than that in the reality because an effect of magnetic fields and 
plasma flows is neglected. Nevertheless, the obtained geometrical scale can be analyzed at least in qualitative (differential) 
terms. We found that the geometrical hight corresponding to $\log \tau = -5$ also varies due to UFs and RPWs and it is 
correlated to the inferred magnetic field variations in the chromosphere. The observed oscillations in the magnetic field
at $\log \tau = -5 $ and corresponding geometrical height suggest that at the onset of a UF magnetic field increases 
and that it is also obtained higher in the atmosphere. There is a hint of a similar scenario in the penumbra, where both 
the magnetic field and its height of inference increases during the relatively hotter phase of RPWs. Increase in the 
magnetic field and the same time in its inferred geometrical height due to oscillations indicate an opposite trend than 
the observed vertical gradient of the magnetic field in the sunspot. So, it is unlikely that the observed temporal variations
in the magnetic field are caused by an opacity effect as suggested by some authors \citep[see,][]{1999ASSL..243..337R, 
2000ApJ...534..989B, 2003A&A...410.1023R, 2003ApJ...588..606K, 2015LRSP...12....6K} to explain the observed 
oscillations in the sunspot's magnetic field in the photosphere.        

\citet{2014ApJ...795....9F} predicted that the magnetic field retrieved from the \ionn{Ca}{ii}\, IR triplet lines produces
pseudo oscillations due to opacity effects. They used magnetohydrodynamic (MHD) simulations of wave propagation 
in a sunspot performed by \citet{2006ApJ...653..739K} and \citet{2010ApJ...719..357F}. \citet{2014ApJ...795....9F} 
found that the weak field approximation applied the \ionn{Ca}{ii}\, IR triplet lines leads to the magnetic field oscillations
with an amplitude of $\sim$100\,G while the MHD simulations do not have such oscillations in the chromosphere.  
\citet{2014ApJ...795....9F} attribute the pseudo oscillations in the magnetic field to change in the height at which the 
\ionn{Ca}{ii}\, IR triple lines are sensitive to the magnetic field: during the quiescent phase, the \ionn{Ca}{ii}\, IR triple
lines are most sensitive to the upper photosphere while as the upward propagating  shock develops the lines become
more sensitive to the chromosphere. 

Our results are computed in hydrostatic equilibrium and we do not perform a full radiation-MHD treatment when computing
the gas pressure scale, but the value of our models is that they are reconstructed from real observations. \citet{2014ApJ...795....9F}
performs MHD simulations of sunspot umbra, but these simulations are also highly simplified in the treatment of heat
 conduction and radiation in the chromosphere.

The latter simulations seem to contradict our observational results, which suggest that the \ionn{Ca}{ii}\,8542\,\AA\, line is
most sensitive to the magnetic field variations at the chromospheric height during UFs as well as the quiescent phase
of the umbra (see, Fig.~\ref{res_funct}). 
Another difference between the prediction of  \citet{2014ApJ...795....9F} and our results is that at the onset of UFs we 
found an increase in the chromospheric magnetic field whereas \citet{2014ApJ...795....9F} suggest a decrease in the magnetic
field during UFs compared to that in the quiescent phase of the umbra. 

Recently, \citet{2017ApJ...845..102H} reported reduced magnetic field in UFs compared to the quiescent umbra which
contradicts our results. However, they also found an enhancement in the magnetic field at the edges of UFs.
Moreover, \citet{2017ApJ...845..102H} observed downflows in UFs and they interpreted these 
downflows in connection with sunspot plumes \citep{1999SoPh..190..437M, 1998ApJ...502L..85B, 1999SoPh..186..141B, 
2001SoPh..198...89B, 2001A&A...368..639F, 2004ApJ...601..546B, 2005ApJ...622.1216B, 2008AnGeo..26.2955D}. However, 
we found that the umbral flashes predominantly show upflows and this might explain the discrepancy between results of 
\citet{2017ApJ...845..102H} and those presented in the current work.     

Very recently, just before the submission of this paper, an independent study has reported very similar results in the outer 
umbra of a sunspot \citep{2018arXiv180300018H} from the inversions of observations acquired in the \ionn{He}{i}\,10830\,\AA\, 
line. They obtain a similar behavior in the outer umbra of a sunspot, although their analysis and inversion model are 
fundamentally different and that \ionn{He}{i} line has an intrinsically different formation mechanism
\citep[see][]{2009ASPC..405..297C,2016A&A...594A.104L} than the \ionn{Ca}{ii}\,8542\,\AA\,
line that we have used in the present study.

Along with the magnetic field, we have analyzed the oscillations in the temperature and LOS velocity in the sunspot. An increase
in the temperature in the atmosphere above $\log\tau = -3$ is observed during UFs compared to that in the quiescent umbra.
At $\log\tau = -5$, we found an average increase of 1\,kK in the temperature due to UFs, which confirms results of 
\citet{2013A&A...556A.115D}. In some of UFs the temperature increases by 1.5\,kK, in addition to the temperature found in the 
quiescent umbra. The average UFs atmosphere shows steep variation in the LOS velocity, where the plasma upflow increases 
from 3\,km\,s$^{-1}$ at $\log\tau = -3$ to 11\,km\,s$^{-1}$ at $\log\tau = -5$. Increase in the temperature and steep gradient 
in the LOS upflow velocity in the atmosphere above $\log\tau = -3$ support shock wave as a mechanism for the formation of
UFs as suggested by \citet{2003A&A...403..277R}, \citet{2006ApJ...640.1153C}, \citet{2010ApJ...722..888B}, \citet{2010ApJ...719..357F},
\citet{2013A&A...556A.115D}, and, \citet{2014ApJ...786..137T}. In the penumbra, we found variations in the temperature at
$\log\tau = -5$ with an amplitude of 0.5\,kK due to RPW.
 
\citet{2018arXiv180205028F} analyzed the effect that a tunable filtergraph instrument can have in inversions, due to the fact that the
observed line profile is not strictly co-temporal at all wavelengths. They find that random errors can appear in the inverted 
parameters if very rapid events occur during the scanning of the instrument. In that case, their inversions could not properly 
reproduce the \emph{observed} profile. Obviously our results can be affected by this source of error, but we want to point 
out that our results are systematic (not random) and we have not found any problems to reproduce the observed profiles 
during UF, even with our inversion setup that uses very few nodes in all physical parameters. So we do not believe that 
these effects are dominating our results.

In summary, we report the observation of oscillations in the chromospheric magnetic field associated to UFs and RPWs in 
the outer umbra and in the penumbra. Our analysis of the magnetic field response function indicates that the observed
oscillations in the sunspot magnetic field are not likely produced by an opacity effect.

\begin{acknowledgements}
This research was supported by the CHROMOBS grant of the Knut och Alice Wallenberg foundation.
JJ is supported by the Research Council of Norway, project 250810, and through its Centres of Excellence scheme,
project number 262622. JdlCR is supported by grants from the Swedish Research Council (2015-03994), the Swedish 
National Space Board (128/15) and the Swedish Civil Contingencies Agency (MSB, 2016-2102). This project has received
funding from the European Research Council (ERC) under the European Union's Horizon 2020 research and innovation 
programme (SUNMAG, grant agreement 759548). The Swedish 1-m Solar Telescope is operated by the Institute for Solar
Physics of Stockholm University in the Spanish Observatorio del Roque de los Muchachos of the Instituto de Astrof\'{\i}sica de Canarias.
Our computations were performed on resources provided by the Swedish National Infrastructure for Computing (SNIC) 
at the PDC Centre for High Performance Computing (PDC-HPC) at the Royal Institute of Technology in Stockholm as well 
as recourses at the High Performance Computing Center North (HPC2N). 
\end{acknowledgements} 

\bibliographystyle{aa} 

\begin{thebibliography}{66}
\expandafter\ifx\csname natexlab\endcsname\relax\def\natexlab#1{#1}\fi

\bibitem[{{Balthasar}(1990)}]{1990SoPh..125...31B}
{Balthasar}, H. 1990, \solphys, 125, 31

\bibitem[{{Balthasar}(2003)}]{2003SoPh..218...85B}
{Balthasar}, H. 2003, \solphys, 218, 85

\bibitem[{{Bard} \& {Carlsson}(2010)}]{2010ApJ...722..888B}
{Bard}, S. \& {Carlsson}, M. 2010, \apj, 722, 888

\bibitem[{{Beckers} \& {Tallant}(1969)}]{1969SoPh....7..351B}
{Beckers}, J.~M. \& {Tallant}, P.~E. 1969, \solphys, 7, 351

\bibitem[{{Bellot Rubio} {et~al.}(2000){Bellot Rubio}, {Collados}, {Ruiz Cobo},
  \& {Rodr{\'{\i}}guez Hidalgo}}]{2000ApJ...534..989B}
{Bellot Rubio}, L.~R., {Collados}, M., {Ruiz Cobo}, B., \& {Rodr{\'{\i}}guez
  Hidalgo}, I. 2000, \apj, 534, 989

\bibitem[{{Bloomfield} {et~al.}(2007){Bloomfield}, {Lagg}, \&
  {Solanki}}]{2007ApJ...671.1005B}
{Bloomfield}, D.~S., {Lagg}, A., \& {Solanki}, S.~K. 2007, \apj, 671, 1005

\bibitem[{{Brosius}(2005)}]{2005ApJ...622.1216B}
{Brosius}, J.~W. 2005, \apj, 622, 1216

\bibitem[{{Brosius} \& {White}(2004)}]{2004ApJ...601..546B}
{Brosius}, J.~W. \& {White}, S.~M. 2004, \apj, 601, 546

\bibitem[{{Brynildsen} {et~al.}(1998){Brynildsen}, {Maltby}, {Brekke},
  {Fredvik}, {Haugan}, {Kjeldseth-Moe}, \& {Wikst{\o}l}}]{1998ApJ...502L..85B}
{Brynildsen}, N., {Maltby}, P., {Brekke}, P., {et~al.} 1998, \apjl, 502, L85

\bibitem[{{Brynildsen} {et~al.}(1999){Brynildsen}, {Maltby}, {Brekke},
  {Haugan}, \& {Kjeldseth-Moe}}]{1999SoPh..186..141B}
{Brynildsen}, N., {Maltby}, P., {Brekke}, P., {Haugan}, S.~V.~H., \&
  {Kjeldseth-Moe}, O. 1999, \solphys, 186, 141

\bibitem[{{Brynildsen} {et~al.}(2001){Brynildsen}, {Maltby}, {Fredvik},
  {Kjeldseth-Moe}, \& {Wilhelm}}]{2001SoPh..198...89B}
{Brynildsen}, N., {Maltby}, P., {Fredvik}, T., {Kjeldseth-Moe}, O., \&
  {Wilhelm}, K. 2001, \solphys, 198, 89

\bibitem[{{Centeno} {et~al.}(2006){Centeno}, {Collados}, \& {Trujillo
  Bueno}}]{2006ApJ...640.1153C}
{Centeno}, R., {Collados}, M., \& {Trujillo Bueno}, J. 2006, \apj, 640, 1153

\bibitem[{{Centeno} {et~al.}(2009){Centeno}, {Trujillo Bueno}, {Uitenbroek}, \&
  {Collados}}]{2009ASPC..405..297C}
{Centeno}, R., {Trujillo Bueno}, J., {Uitenbroek}, H., \& {Collados}, M. 2009,
  in Astronomical Society of the Pacific Conference Series, Vol. 405, Solar
  Polarization 5: In Honor of Jan Stenflo, ed. S.~V. {Berdyugina}, K.~N.
  {Nagendra}, \& R.~{Ramelli}, 297

\bibitem[{{Dammasch} {et~al.}(2008){Dammasch}, {Curdt}, {Dwivedi}, \&
  {Parenti}}]{2008AnGeo..26.2955D}
{Dammasch}, I.~E., {Curdt}, W., {Dwivedi}, B.~N., \& {Parenti}, S. 2008,
  Annales Geophysicae, 26, 2955

\bibitem[{{de la Cruz Rodr{\'{\i}}guez} {et~al.}(2015){de la Cruz
  Rodr{\'{\i}}guez}, {L{\"o}fdahl}, {S{\"u}tterlin}, {Hillberg}, \& {Rouppe van
  der Voort}}]{2015A&A...573A..40D}
{de la Cruz Rodr{\'{\i}}guez}, J., {L{\"o}fdahl}, M.~G., {S{\"u}tterlin}, P.,
  {Hillberg}, T., \& {Rouppe van der Voort}, L. 2015, \aap, 573, A40

\bibitem[{{de la Cruz Rodr{\'{\i}}guez} \&
  {Piskunov}(2013)}]{2013ApJ...764...33D}
{de la Cruz Rodr{\'{\i}}guez}, J. \& {Piskunov}, N. 2013, \apj, 764, 33

\bibitem[{{de la Cruz Rodr{\'{\i}}guez} {et~al.}(2013){de la Cruz
  Rodr{\'{\i}}guez}, {Rouppe van der Voort}, {Socas-Navarro}, \& {van
  Noort}}]{2013A&A...556A.115D}
{de la Cruz Rodr{\'{\i}}guez}, J., {Rouppe van der Voort}, L., {Socas-Navarro},
  H., \& {van Noort}, M. 2013, \aap, 556, A115

\bibitem[{{De Moortel} {et~al.}(2002){De Moortel}, {Ireland}, {Hood}, \&
  {Walsh}}]{2002A&A...387L..13D}
{De Moortel}, I., {Ireland}, J., {Hood}, A.~W., \& {Walsh}, R.~W. 2002, \aap,
  387, L13

\bibitem[{{Felipe} {et~al.}(2010){Felipe}, {Khomenko}, \&
  {Collados}}]{2010ApJ...719..357F}
{Felipe}, T., {Khomenko}, E., \& {Collados}, M. 2010, \apj, 719, 357

\bibitem[{{Felipe} {et~al.}(2014){Felipe}, {Socas-Navarro}, \&
  {Khomenko}}]{2014ApJ...795....9F}
{Felipe}, T., {Socas-Navarro}, H., \& {Khomenko}, E. 2014, \apj, 795, 9

\bibitem[{{Felipe} {et~al.}(2018){Felipe}, {Socas-Navarro}, \&
  {Przybylski}}]{2018arXiv180205028F}
{Felipe}, T., {Socas-Navarro}, H., \& {Przybylski}, D. 2018, ArXiv e-prints [arXiv:1802.05028]

\bibitem[{{Fludra}(2001)}]{2001A&A...368..639F}
{Fludra}, A. 2001, \aap, 368, 639

\bibitem[{{Fontenla} {et~al.}(1993){Fontenla}, {Avrett}, \&
  {Loeser}}]{1993ApJ...406..319F}
{Fontenla}, J.~M., {Avrett}, E.~H., \& {Loeser}, R. 1993, \apj, 406, 319

\bibitem[{{Gurman}(1987)}]{1987SoPh..108...61G}
{Gurman}, J.~B. 1987, \solphys, 108, 61

\bibitem[{{Gurman} {et~al.}(1982){Gurman}, {Leibacher}, {Shine}, {Woodgate}, \&
  {Henze}}]{1982ApJ...253..939G}
{Gurman}, J.~B., {Leibacher}, J.~W., {Shine}, R.~A., {Woodgate}, B.~E., \&
  {Henze}, W. 1982, \apj, 253, 939

\bibitem[{{Henriques}(2012)}]{2012A&A...548A.114H}
{Henriques}, V.~M.~J. 2012, \aap, 548, A114

\bibitem[{{Henriques} {et~al.}(2017){Henriques}, {Mathioudakis},
  {Socas-Navarro}, \& {de la Cruz Rodr{\'{\i}}guez}}]{2017ApJ...845..102H}
{Henriques}, V.~M.~J., {Mathioudakis}, M., {Socas-Navarro}, H., \& {de la Cruz
  Rodr{\'{\i}}guez}, J. 2017, \apj, 845, 102

\bibitem[{{Horn} {et~al.}(1997){Horn}, {Staude}, \&
  {Landgraf}}]{1997SoPh..172...69H}
{Horn}, T., {Staude}, J., \& {Landgraf}, V. 1997, \solphys, 172, 69

\bibitem[{{Houston} {et~al.}(2018){Houston}, {Jess}, {Asensio Ramos}, {Grant},
  {Beck}, {Norton}, \& {Krishna Prasad}}]{2018arXiv180300018H}
{Houston}, S.~J., {Jess}, D.~B., {Asensio Ramos}, A., {et~al.} 2018, ArXiv e-prints [arXiv:1803.00018]

\bibitem[{{Janett} {et~al.}(2017){Janett}, {Carlin}, {Steiner}, \&
  {Belluzzi}}]{2017ApJ...840..107J}
{Janett}, G., {Carlin}, E.~S., {Steiner}, O., \& {Belluzzi}, L. 2017, \apj,
  840, 107

\bibitem[{{Joshi} {et~al.}(2017{\natexlab{a}}){Joshi}, {Lagg}, {Hirzberger}, \&
  {Solanki}}]{2017A&A...604A..98J}
{Joshi}, J., {Lagg}, A., {Hirzberger}, J., \& {Solanki}, S.~K.
  2017{\natexlab{a}}, \aap, 604, A98

\bibitem[{{Joshi} {et~al.}(2017{\natexlab{b}}){Joshi}, {Lagg}, {Hirzberger},
  {Solanki}, \& {Tiwari}}]{2017A&A...599A..35J}
{Joshi}, J., {Lagg}, A., {Hirzberger}, J., {Solanki}, S.~K., \& {Tiwari}, S.~K.
  2017{\natexlab{b}}, \aap, 599, A35

\bibitem[{{Joshi} {et~al.}(2016){Joshi}, {Lagg}, {Solanki}, {Feller},
  {Collados}, {Orozco Su{\'a}rez}, {Schlichenmaier}, {Franz}, {Balthasar},
  {Denker}, {Berkefeld}, {Hofmann}, {Kiess}, {Nicklas}, {Pastor Yabar},
  {Rezaei}, {Schmidt}, {Schmidt}, {Sobotka}, {Soltau}, {Staude}, {Strassmeier},
  {Volkmer}, {von der L{\"u}he}, \& {Waldmann}}]{2016A&A...596A...8J}
{Joshi}, J., {Lagg}, A., {Solanki}, S.~K., {et~al.} 2016, \aap, 596, A8

\bibitem[{{Kallunki} \& {Riehokainen}(2012)}]{2012SoPh..280..347K}
{Kallunki}, J. \& {Riehokainen}, A. 2012, \solphys, 280, 347

\bibitem[{{Kanoh} {et~al.}(2016){Kanoh}, {Shimizu}, \&
  {Imada}}]{2016ApJ...831...24K}
{Kanoh}, R., {Shimizu}, T., \& {Imada}, S. 2016, \apj, 831, 24

\bibitem[{{Kentischer} \& {Mattig}(1995)}]{1995A&A...300..539K}
{Kentischer}, T.~J. \& {Mattig}, W. 1995, \aap, 300, 539

\bibitem[{{Khomenko} \& {Collados}(2006)}]{2006ApJ...653..739K}
{Khomenko}, E. \& {Collados}, M. 2006, \apj, 653, 739

\bibitem[{{Khomenko} \& {Collados}(2015)}]{2015LRSP...12....6K}
{Khomenko}, E. \& {Collados}, M. 2015, Living Reviews in Solar Physics, 12, 6

\bibitem[{{Khomenko} {et~al.}(2003){Khomenko}, {Collados}, \& {Bellot
  Rubio}}]{2003ApJ...588..606K}
{Khomenko}, E.~V., {Collados}, M., \& {Bellot Rubio}, L.~R. 2003, \apj, 588,
  606

\bibitem[{{Kupke} {et~al.}(2000){Kupke}, {Labonte}, \&
  {Mickey}}]{2000SoPh..191...97K}
{Kupke}, R., {Labonte}, B.~J., \& {Mickey}, D.~L. 2000, \solphys, 191, 97

\bibitem[{{Landgraf}(1997)}]{1997AN....318..129L}
{Landgraf}, V. 1997, Astronomische Nachrichten, 318, 129

\bibitem[{{Leenaarts} {et~al.}(2014){Leenaarts}, {de la Cruz Rodr{\'{\i}}guez},
  {Kochukhov}, \& {Carlsson}}]{2014ApJ...784L..17L}
{Leenaarts}, J., {de la Cruz Rodr{\'{\i}}guez}, J., {Kochukhov}, O., \&
  {Carlsson}, M. 2014, \apjl, 784, L17

\bibitem[{{Leenaarts} {et~al.}(2016){Leenaarts}, {Golding}, {Carlsson},
  {Libbrecht}, \& {Joshi}}]{2016A&A...594A.104L}
{Leenaarts}, J., {Golding}, T., {Carlsson}, M., {Libbrecht}, T., \& {Joshi}, J.
  2016, \aap, 594, A104

\bibitem[{{Lites}(1984)}]{1984ApJ...277..874L}
{Lites}, B.~W. 1984, \apj, 277, 874

\bibitem[{{Lites}(1986)}]{1986ApJ...301..992L}
{Lites}, B.~W. 1986, \apj, 301, 992

\bibitem[{{Lites} {et~al.}(1998){Lites}, {Thomas}, {Bogdan}, \&
  {Cally}}]{1998ApJ...497..464L}
{Lites}, B.~W., {Thomas}, J.~H., {Bogdan}, T.~J., \& {Cally}, P.~S. 1998, \apj,
  497, 464

\bibitem[{{Maltby} {et~al.}(1999){Maltby}, {Brynildsen}, {Fredvik},
  {Kjeldseth-Moe}, \& {Wilhelm}}]{1999SoPh..190..437M}
{Maltby}, P., {Brynildsen}, N., {Fredvik}, T., {Kjeldseth-Moe}, O., \&
  {Wilhelm}, K. 1999, \solphys, 190, 437

\bibitem[{{Orozco Suarez} {et~al.}(2005){Orozco Suarez}, {Lagg}, \&
  {Solanki}}]{2005ESASP.596E..59O}
{Orozco Suarez}, D., {Lagg}, A., \& {Solanki}, S.~K. 2005, in ESA Special
  Publication, Vol. 596, Chromospheric and Coronal Magnetic Fields, ed. D.~E.
  {Innes}, A.~{Lagg}, \& S.~A. {Solanki}, 59.1

\bibitem[{{Quintero Noda} {et~al.}(2016){Quintero Noda}, {Shimizu}, {de la Cruz
  Rodr{\'{\i}}guez}, {Katsukawa}, {Ichimoto}, {Anan}, \&
  {Suematsu}}]{2016MNRAS.459.3363Q}
{Quintero Noda}, C., {Shimizu}, T., {de la Cruz Rodr{\'{\i}}guez}, J., {et~al.}
  2016, \mnras, 459, 3363

\bibitem[{{Rouppe van der Voort} {et~al.}(2003){Rouppe van der Voort},
  {Rutten}, {S{\"u}tterlin}, {Sloover}, \& {Krijger}}]{2003A&A...403..277R}
{Rouppe van der Voort}, L.~H.~M., {Rutten}, R.~J., {S{\"u}tterlin}, P.,
  {Sloover}, P.~J., \& {Krijger}, J.~M. 2003, \aap, 403, 277

\bibitem[{{R{\"u}edi} \& {Cally}(2003)}]{2003A&A...410.1023R}
{R{\"u}edi}, I. \& {Cally}, P.~S. 2003, \aap, 410, 1023

\bibitem[{{R{\"u}edi} {et~al.}(1999){R{\"u}edi}, {Solanki}, {Bogdan}, \&
  {Cally}}]{1999ASSL..243..337R}
{R{\"u}edi}, I., {Solanki}, S.~K., {Bogdan}, T., \& {Cally}, P. 1999, in
  Astrophysics and Space Science Library, Vol. 243, Polarization, ed. K.~N.
  {Nagendra} \& J.~O. {Stenflo}, 337--347

\bibitem[{{Rueedi} {et~al.}(1995){Rueedi}, {Solanki}, \&
  {Livingston}}]{1995A&A...302..543R}
{Rueedi}, I., {Solanki}, S.~K., \& {Livingston}, W. 1995, \aap, 302, 543

\bibitem[{{Rueedi} {et~al.}(1998){Rueedi}, {Solanki}, {Stenflo}, {Tarbell}, \&
  {Scherrer}}]{1998A&A...335L..97R}
{Rueedi}, I., {Solanki}, S.~K., {Stenflo}, J.~O., {Tarbell}, T., \& {Scherrer},
  P.~H. 1998, \aap, 335, L97

\bibitem[{{Schad} {et~al.}(2015){Schad}, {Penn}, {Lin}, \&
  {Tritschler}}]{2015SoPh..290.1607S}
{Schad}, T.~A., {Penn}, M.~J., {Lin}, H., \& {Tritschler}, A. 2015, \solphys,
  290, 1607

\bibitem[{{Scharmer}(2006)}]{2006A&A...447.1111S}
{Scharmer}, G.~B. 2006, \aap, 447, 1111

\bibitem[{{Scharmer} {et~al.}(2003){Scharmer}, {Bjelksjo}, {Korhonen},
  {Lindberg}, \& {Petterson}}]{2003SPIE.4853..341S}
{Scharmer}, G.~B., {Bjelksjo}, K., {Korhonen}, T.~K., {Lindberg}, B., \&
  {Petterson}, B. 2003, in \procspie, Vol. 4853, Innovative Telescopes and
  Instrumentation for Solar Astrophysics, ed. S.~L. {Keil} \& S.~V. {Avakyan},
  341--350

\bibitem[{{Scharmer} {et~al.}(2010){Scharmer}, {L{\"o}fdahl}, {van Werkhoven},
  \& {de la Cruz Rodr{\'{\i}}guez}}]{2010A&A...521A..68S}
{Scharmer}, G.~B., {L{\"o}fdahl}, M.~G., {van Werkhoven}, T.~I.~M., \& {de la
  Cruz Rodr{\'{\i}}guez}, J. 2010, \aap, 521, A68

\bibitem[{{Scharmer} {et~al.}(2008){Scharmer}, {Narayan}, {Hillberg}, {de la
  Cruz Rodriguez}, {L{\"o}fdahl}, {Kiselman}, {S{\"u}tterlin}, {van Noort}, \&
  {Lagg}}]{2008ApJ...689L..69S}
{Scharmer}, G.~B., {Narayan}, G., {Hillberg}, T., {et~al.} 2008, \apjl, 689,
  L69

\bibitem[{{Schnerr} {et~al.}(2011){Schnerr}, {de La Cruz Rodr{\'{\i}}guez}, \&
  {van Noort}}]{2011A&A...534A..45S}
{Schnerr}, R.~S., {de La Cruz Rodr{\'{\i}}guez}, J., \& {van Noort}, M. 2011,
  \aap, 534, A45

\bibitem[{{Socas-Navarro} {et~al.}(2015){Socas-Navarro}, {de la Cruz
  Rodr{\'{\i}}guez}, {Asensio Ramos}, {Trujillo Bueno}, \& {Ruiz
  Cobo}}]{2015A&A...577A...7S}
{Socas-Navarro}, H., {de la Cruz Rodr{\'{\i}}guez}, J., {Asensio Ramos}, A.,
  {Trujillo Bueno}, J., \& {Ruiz Cobo}, B. 2015, \aap, 577, A7

\bibitem[{{Socas-Navarro} {et~al.}(2000){Socas-Navarro}, {Trujillo Bueno}, \&
  {Ruiz Cobo}}]{2000Sci...288.1396S}
{Socas-Navarro}, H., {Trujillo Bueno}, J., \& {Ruiz Cobo}, B. 2000, Science,
  288, 1396

\bibitem[{{Thomas} {et~al.}(1987){Thomas}, {Lites}, {Gurman}, \&
  {Ladd}}]{1987ApJ...312..457T}
{Thomas}, J.~H., {Lites}, B.~W., {Gurman}, J.~B., \& {Ladd}, E.~F. 1987, \apj,
  312, 457

\bibitem[{{Tian} {et~al.}(2014){Tian}, {DeLuca}, {Reeves}, {McKillop}, {De
  Pontieu}, {Mart{\'{\i}}nez-Sykora}, {Carlsson}, {Hansteen}, {Kleint},
  {Cheung}, {Golub}, {Saar}, {Testa}, {Weber}, {Lemen}, {Title}, {Boerner},
  {Hurlburt}, {Tarbell}, {Wuelser}, {Kankelborg}, {Jaeggli}, \&
  {McIntosh}}]{2014ApJ...786..137T}
{Tian}, H., {DeLuca}, E., {Reeves}, K.~K., {et~al.} 2014, \apj, 786, 137

\bibitem[{{van Noort} {et~al.}(2005){van Noort}, {Rouppe van der Voort}, \&
  {L{\"o}fdahl}}]{2005SoPh..228..191V}
{van Noort}, M., {Rouppe van der Voort}, L., \& {L{\"o}fdahl}, M.~G. 2005,
  \solphys, 228, 191

\bibitem[{{Wittmann}(1969)}]{1969SoPh....7..366W}
{Wittmann}, A. 1969, \solphys, 7, 366

\end{thebibliography}

\begin{appendix}
\section{Quality of the fits}\label{ap:fit}
\begin{figure*}
\centering
       \includegraphics[width = 0.48\textwidth]{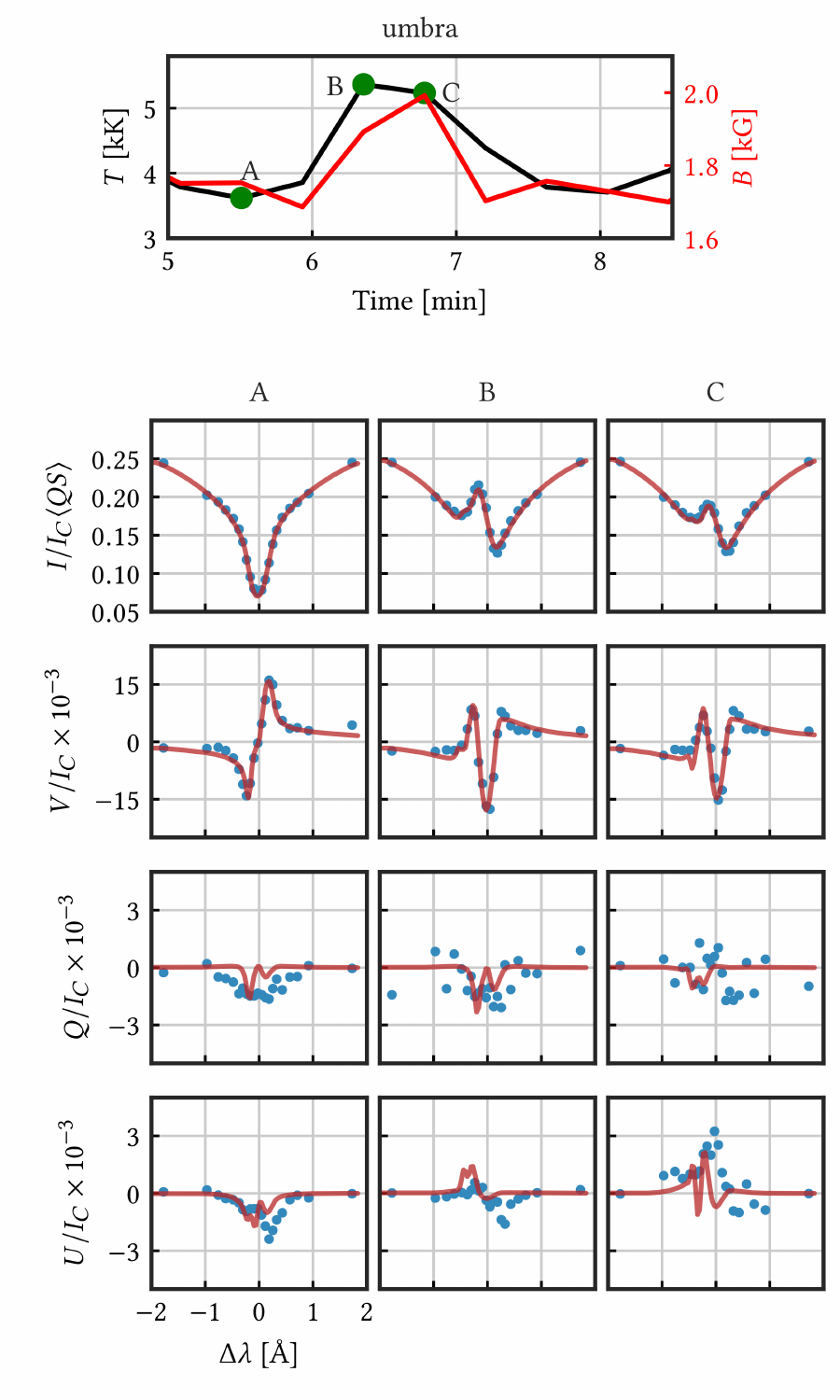}       \includegraphics[width = 0.48\textwidth]{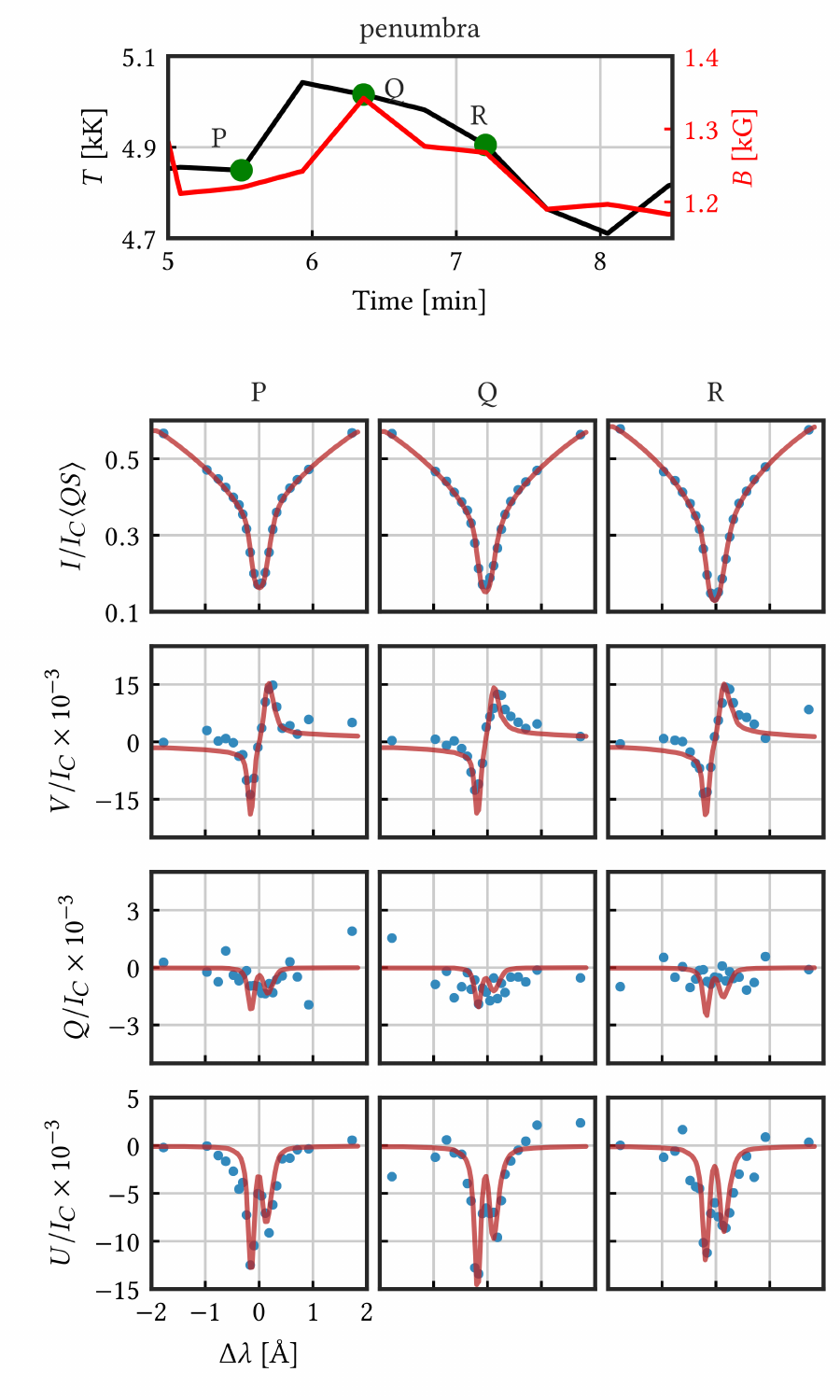}

       \caption{Examples of observed and best fitted profiles in the \ionn{Ca}{ii}\,8542\,\AA\, 
                      line in the umbra (right) and penumbra (left). For each set, the top panel shows $T$ (black) and $B$ (red) at
                      $\log \tau = -5$. The observed (blue dots) and best fitted (red curves) Stokes $I$, 
                      $V$, $Q$, and $U$ profiles are represented in each column for 3 different time stamps, which are indicated with capital letters. }
                \label{stokes_umbra}
\end{figure*}

We have
assessed the match between the observed and synthesized Stokes profiles from the inverted model atmosphere. We show some
examples of the observed and best fitted Stokes profiles in the \ionn{Ca}{ii}\,8542\,\AA\, line in Fig.~\ref{stokes_umbra}.
The top left panel in Fig.~\ref{stokes_umbra} represents the variation in 
$T$ and $B$  at $\log\tau = -5$ during a passage of a UFs 
at $r = 13.\arcsec8$. The observed and fitted Stokes profiles at different phases of the UF 
are depicted in the lower panels marked as 'A', 'B' and 'C' in Fig.~\ref{stokes_umbra}. 
At location 'B', where the $T$ is enhanced due to the UF, Stokes $I$ profile shows a blue shifted line core in emission.  
The match between the synthesized and observed Stokes $I$ and $V$ profiles are very good at all three locations.
In the umbra, the Stokes Q \&\ U signal is below the noise level. 
Therefore, hereafter we will not discuss the inferred transverse component of the umbral magnetic field.
 
The top right panel in Fig.~\ref{stokes_umbra} displays the variations of $T$  
and $B$ at $\log\tau = -5$ during a passage of a RPW at $r = 7.\arcsec4$. Similar to the umbra, we plotted the observed and fitted Stokes profiles at different phases (marked as 'P', 'Q' and, 'R' )
of the RPW. In the penumbra, Stokes $I$, $V$ and $U$ profiles are fitted very well, but not Stokes $Q$ profiles. The 
Stokes $Q$ profiles are weak and very noisy, and this is purely due to the choice of the 
$red$ cut shown in Fig.~\ref{context} and the orientation of the sunspot's magnetic field at that location
in the penumbra. In other parts of the sunspot penumbra, Stokes $Q$ profiles have a significant signal. The Stokes $U$
profiles have as much as $\sim0.013I_C$ signal and they are very well fitted by the inversions which gives a somewhat reliable 
measurement of $B_t$. 

\section{Sensitivity to the magnetic field vector}\label{ap:sen}

\subsection{Response functions during UFs and quiescence}\label{sec:rf}

\begin{figure*}
\centering
       \includegraphics[width = 0.92\textwidth]{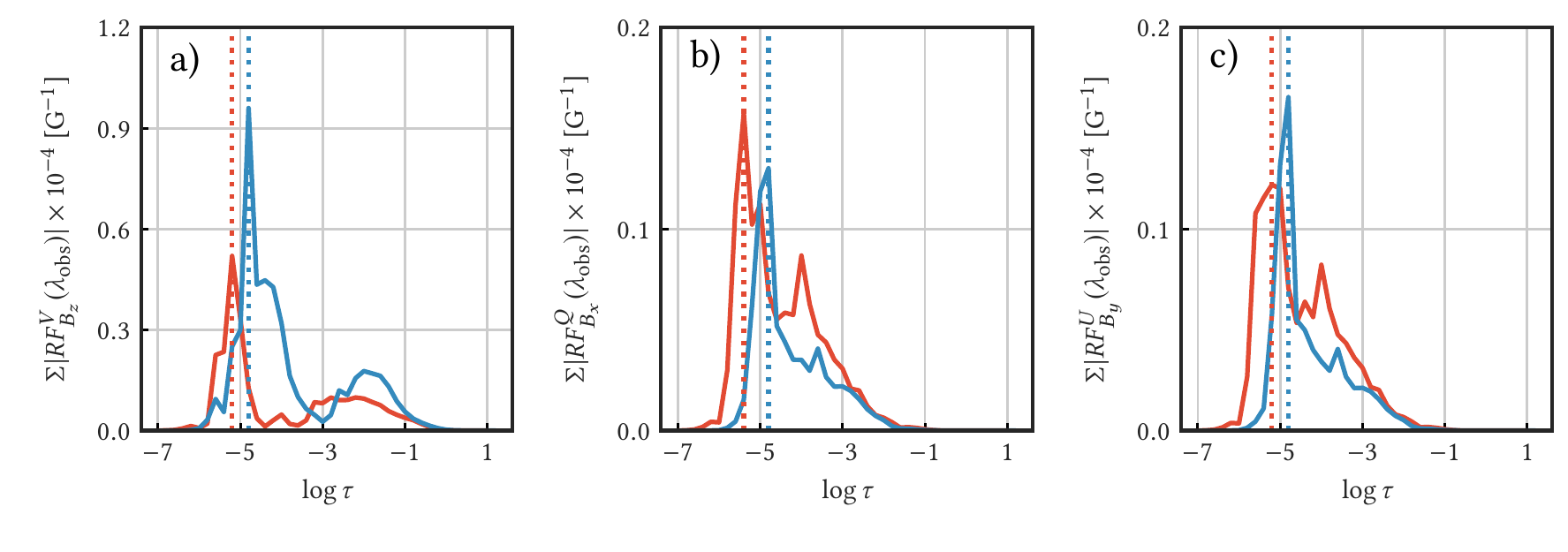}
       \caption{Wavelength integrated magnetic field response functions of the \ionn{Ca}{ii}\,8542\,\AA\, line. Panel \textbf{a)} 
                shows sum over observed wavelengths of the absolute value response function of Stokes $V$ to $B_z$, 
                $\sum \mid \rm{RF}_{B_z}^V (\lambda_{\rm{obs}}) \mid$. Panels \textbf{b)} and \textbf{c)} display
                $\sum \mid \rm{RF}_{B_x}^Q (\lambda_{\rm{obs}}) \mid$ and $\sum \mid \rm{RF}_{B_y}^U (\lambda_{\rm{obs}}) \mid$.
                The \textit{red} curves correspond to an UF atmosphere and \textit{blue} curves indicate a 
                quiescent umbra atmosphere. The maximum of each curve is indicated with a vertical dotted line with the same color coding.}
       \label{res_funct}        
\end{figure*}

\begin{figure}
\centering
       \includegraphics[width = \columnwidth]{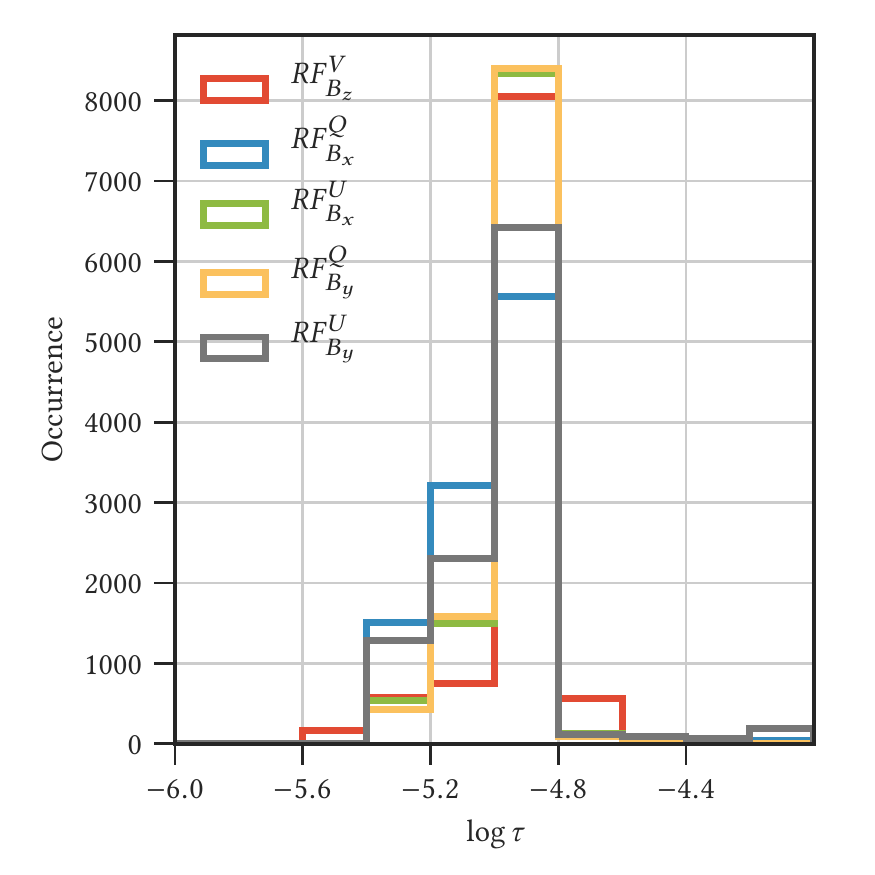}
       \caption{Distribution of $\log \tau$ values corresponding to the maximum value of 
        $\sum \mid \rm{RF}_{B_z}^V (\lambda_{\rm{obs}}) \mid$ (\textit{red}),
        $\sum \mid \rm{RF}_{B_x}^Q (\lambda_{\rm{obs}}) \mid$ (\textit{blue}),
        $\sum \mid \rm{RF}_{B_x}^U (\lambda_{\rm{obs}}) \mid$ (\textit{green}),
        $\sum \mid \rm{RF}_{B_y}^Q (\lambda_{\rm{obs}}) \mid$ (\textit{yellow}) and,
        $\sum \mid \rm{RF}_{B_y}^U (\lambda_{\rm{obs}}) \mid$ (\textit{gray})
        for the \ionn{Ca}{ii}\,8542\,\AA\, line.
        These histograms are calculated within the spatial interval $r =  2\varcsec$ - $16\varcsec$.}
                
        \label{histo}
\end{figure}

Using the resulting models from our inversions, we have calculated $RF_{\rm{B_i}}^{\rm{S}}$  for the \ionn{Ca}{ii}\,8542\,\AA\, line in every pixel for all
time steps. Here $B_{\rm{i}}$ denotes three components of the magnetic field and $S$
represent Stokes $Q$, $U$ and, $V$. Fig.~\ref{res_funct} shows the sum of the absolute 
value of the $RF$ over wavelength ($\sum_\lambda \mid RF(\lambda_{\rm{obs}})\mid$). Our results indicate that during UFs, the 
maximum response of the Stokes $Q$, $U$ and, $V$ profiles are located at lower $\log \tau$ values compared to those 
during the quiescent phase in the umbra. We found $\log \tau$ value corresponding to 
the maximum of $\sum \mid RF(\lambda_{\rm{obs}})\mid$  for each pixel within the sunspot 
in our space-time map and histograms of obtained $\log \tau$ values are plotted in Fig.~\ref{histo}.

Our results also indicate that for these models,
the \ionn{Ca}{ii}\,8542\,\AA\, line has maximum response to the magnetic field between $\log\tau = -4.8$ and 
$\log\tau = -5.2$.
These results also suggest that inversions assuming a constant magnetic
field will greatly weight this range of the atmosphere.

\subsection{The effect of gradients in the inversions of $B$}

\begin{figure}
\centering
        \includegraphics[width = \columnwidth]{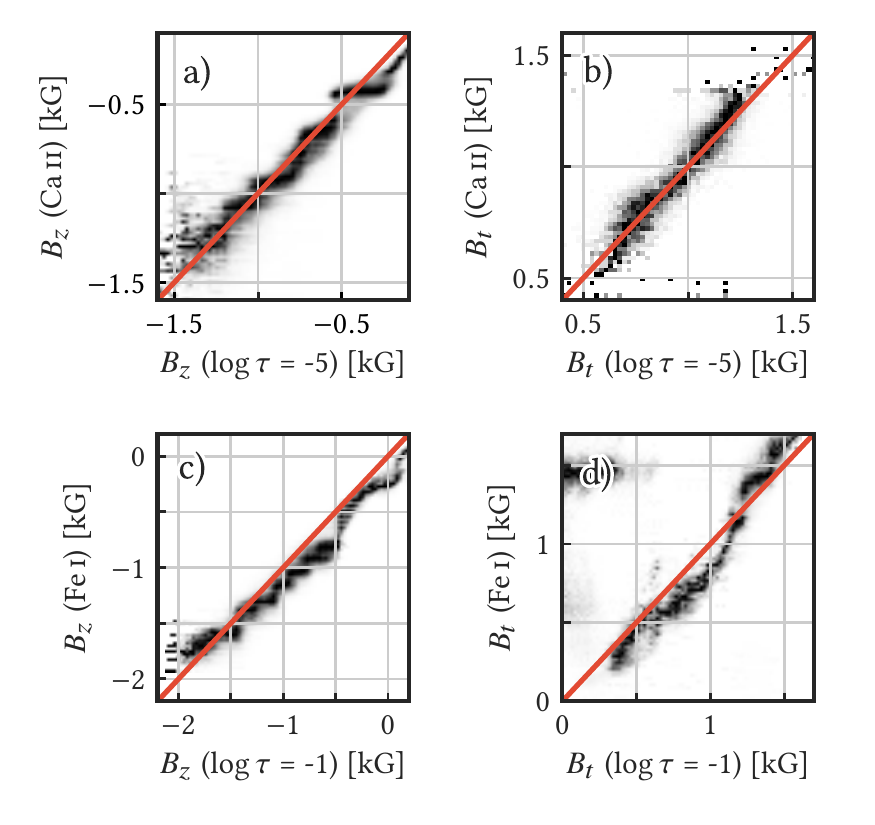}
       \caption{Comparison of the magnetic field retrieved from the inversions with a single node in 
        each component of the magnetic field vs that from the inversions that includes a linear gradient in the magnetic field. 
        $B_{\rm{(z,t)}}(\log = -5, -1)$ is taken from the inversions of the \ionn{Ca}{ii}\,8542\,\AA\, line including magnetic field gradient. 
        The depth-independent magnetic field values derived from the inversions of the \ionn{Ca}{ii}\,\,8542\,\AA\, 
        line and the \ionn{Fe}{i}\,6301.5\,\&\,6302.5\,\AA\,  lines are indicated as 
       $B_{\rm{(z,t)}}$(\ionn{Ca}{ii}) and $B_{\rm{(z,t)}}$(\ionn{Fe}{i}), respectively.
       \textbf{a)}: Normalized two-dimensional histogram of $B_{z}(\log \tau = -5)$ vs.
       $B_{z}$(\ionn{Ca}{ii}). \textbf{b)}: similar to panel \textbf{a)}, but it compares $B_{t}(\log \tau = -5)$ and $B_{t}$(\ionn{Ca}{ii}).
      \textbf{c)}: Normalized two-dimensional histogram of $B_{z}(\log \tau = -1)$ vs. 
       $B_{z}$(\ionn{Fe}{i}). \textbf{d)}: similar to panel \textbf{c)}, but it compares $B_{t}(\log \tau = -1)$ and $B_{t}$(\ionn{Fe}{i}).
      The \textit{red} line in all the panels illustrates a one to one correspondence.
       }
        \label{comp}
\end{figure}

We have compared the magnetic field properties obtained from the inversion of the \ionn{Ca}{ii}\,8542\,\AA\, line with 
the assumption that the magnetic field varies linearly with $\log \tau$, to those retrieved by assuming that the 
magnetic field is constant. The latter comparison is illustrated in Fig.~\ref{comp} for the vertical and horizontal 
components of the magnetic field. In Fig.~\ref{comp}, $B_{\rm{(z,t)}}(\log \tau = -5, -1 )$ represents the magnetic 
field retrieved through the inversions of the \ionn{Ca}{ii}\,8542\,\AA\, line under the assumption that the magnetic field varies 
linearly with $\log \tau$. Whereas, $B_{\rm{(z,t)}}$(\ionn{Ca}{ii}) and $B_{\rm{(z,t)}}$(\ionn{Fe}{i}) 
corresponds to the magnetic field obtained from the inversions of the \ionn{Ca}{ii}\,8542\,\AA\, line and  
the \ionn{Fe}{i}\,6301.5\,\&\,6302.5\,\AA\, lines, assuming a depth-independent magnetic field.
As expected from our tests in \S\ref{sec:rf}, even when we allow for
gradients in the magnetic field stratification, the retrieved values around the maximum of the $RF$, $\log \tau = -5$, greatly correlate with
the results from the depth-independent magnetic field inversions. The Pearson coefficients, $p$, between $B_{\rm{z}}(\log\tau= -5)$ and $B_{\rm{z}}$(\ionn{Ca}{ii}) and between $B_{\rm{t}}(\log\tau= -5)$ and $B_{\rm{t}}$(\ionn{Ca}{ii}) are 0.83 and 0.81  (see panel \textbf{a)} and \textbf{b)}), respectively.

Based on previous studies, the \ionn{Fe}{i}\,6301.5\,\&\,6302.5\,\AA\, lines are expected to have maximum response to the magnetic field at 
$\log \tau \approx-1$ \citep[e.g., see Fig.~9 of][among others]{2017A&A...599A..35J}. Our results for $B_{\rm{z}}(\log \tau = -1)$ from the inversion of 
the \ionn{Ca}{ii}\,8542\,\AA\, line are very well correlated ($p=0.83$, panel \textbf{c)}) with those from single node inversions of the \ionn{Fe}{i}\,6301.5\,\&\,6302.5\,\AA\, lines. The correlation for the transverse component is worse ($p=0.63$, panel \textbf{d)}) but the weaker Stokes $Q$ and $U$ signals may not encode sufficient information to accurately retrieve gradients in the horizontal component in our observations.

We note that although the \ionn{Ca}{ii}\,8542\,\AA\, line is not expected to have a strong response to photospheric magnetic fields, 
our inversions with gradients in this line yield similar values of $B_z$ than those from the inversions of the \ionn{Fe}{i}\,6301.5\,\&\,6302.5\,\AA\, lines, 
and there is clear Stokes~$V$ signal present in the wings of the former line.

\end{appendix}

\end{document}